%% file: main.tex
\documentclass[acmsmall,screen,nonacm]{acmart} 

\usepackage{multicol}
\usepackage{listings}
\usepackage{subcaption}
\usepackage{xspace}
\usepackage{xcolor}
\usepackage{xparse}
\usepackage{stmaryrd}
\usepackage{tikz-cd}
\usepackage{cleveref}
\usepackage{lstlinebgrd}
\usepackage{algorithm}
\usepackage{algpseudocode}
\usepackage{algorithmicx}
\usepackage{wrapfig}
\usepackage{dirtree}

\usepackage{tikz}
\usetikzlibrary{arrows.meta, shapes.geometric, positioning, matrix, calc, fit, backgrounds, bending}
\pgfdeclarelayer{background0}
\pgfdeclarelayer{background1}
\pgfdeclarelayer{background2}
\pgfsetlayers{background2,background1,background0,main}

\input{macros} 

\begin{document}

\title{A Compilation Model for Divergent Parallel Workloads}


\author{Alexander J Root}
\orcid{0000-0001-6221-1389}
\affiliation{%
  \institution{Stanford University}
  \streetaddress{353 Jane Stanford Way}
  \city{Stanford}
  \state{CA}
  \country{USA}
  \postcode{94305}
}
\email{ajroot@cs.stanford.edu}

\author{Christophe Gyurgyik}
\orcid{0000-0001-8493-1133}
\affiliation{%
  \institution{Stanford University}
  \streetaddress{353 Jane Stanford Way}
  \city{Stanford}
  \state{CA}
  \country{USA}
  \postcode{94305}
}
\email{cpg@cs.stanford.edu}

\author{Devanshu Ladsaria}
\orcid{0009-0002-5598-8231}
\affiliation{%
  \institution{Stanford University}
  \streetaddress{353 Jane Stanford Way}
  \city{Stanford}
  \state{CA}
  \country{USA}
  \postcode{94305}
}
\email{devanshu@cs.stanford.edu}


\author{Rohan Sawhney}
\orcid{0000-0002-3661-1554}
\affiliation{%
  \institution{NVIDIA}
  \city{Santa Clara}
  \state{CA}
  \country{USA}
  \postcode{95051}
}
\email{rsawhney@nvidia.com}

\author{Jonathan Ragan-Kelley}
\orcid{0000-0001-6243-9543}
\affiliation{%
  \institution{MIT}
  \city{Cambridge}
  \state{MA}
  \country{USA}
  \postcode{02139}
}
\email{jrk@mit.edu}

\author{Andrew Adams}
\orcid{0000-0002-6766-670X}
\affiliation{%
  \institution{Adobe}
  \city{San Francisco}
  \state{CA}
  \country{USA}
  \postcode{94103}
}
\email{andrew.b.adams@gmail.com}

\author{Matt Pharr}
\orcid{0000-0002-0566-8291}
\affiliation{%
  \institution{NVIDIA}
  \city{Santa Clara}
  \state{CA}
  \country{USA}
  \postcode{95051}
}
\email{matt@pharr.org}

\author{Kayvon Fatahalian}
\orcid{0000-0001-8754-0429}
\affiliation{%
  \institution{Stanford University}
  \streetaddress{353 Jane Stanford Way}
  \city{Stanford}
  \state{CA}
  \country{USA}
  \postcode{94305}
}
\email{kayvonf@cs.stanford.edu}

\author{Fredrik Kjolstad}
\orcid{0000-0002-2267-903X}
\affiliation{%
  \institution{Stanford University}
  \streetaddress{353 Jane Stanford Way}
  \city{Stanford}
  \state{CA}
  \country{USA}
  \postcode{94305}
}
\email{kjolstad@cs.stanford.edu}

\renewcommand{\shortauthors}{Root et al.}

\begin{abstract}

Efficient execution on modern parallel machines requires more than abundant parallelism; it demands instruction coherence, memory coherence, and load balance to maximize SIMD utilization and memory bandwidth.
Efficiently mapping \textit{divergent} parallel workloads to modern machines, where control flow, memory accesses, and unbounded iteration depend on per-thread runtime values, often requires global program restructuring that trades off increased coherence with reduced locality, parallelism, or extra work. 
We observe that the key optimizations used to accelerate divergent parallel programs can be expressed as compositions of well-known program transformations from the functional programming and loop optimization communities. 
We operationalize this insight through \textit{parallel dispatch graphs}: a compilation model that represents divergent programs by representing control flow, data flow between logical stages, and parallel loops within a single abstraction.
To facilitate rapid exploration of points in the optimization space, we provide a scheduling language that includes composable transformations of data-dependent control flow and data flow, together with standard loop transformations and hardware bindings.
This separation of concerns allows the core algorithm to remain portable, while the execution schedule can be specialized to the specifics of target hardware and input data values.
We evaluate our system using popular applications from the domain of spatial computing, and show that the optimization strategies of hand-optimized parallel tree traversals, path tracers, and Monte Carlo PDE solvers can be reproduced as schedules in our system.
Our generated code matches or outperforms the hand-optimized implementations (geomeans $1.02$--$2.95\times$ faster).
Additionally, we demonstrate empirically that the best schedule varies across programs, machines, and inputs.
Through manual exploration we find per-scene path tracing schedules that beat the expert-written schedules by a further $1.3\times$ on the CPU and $1.6\times$ on the GPU (geomean), up to $13\times$.

\end{abstract}

\maketitle 

\section{Introduction}

The Single Program, Multiple Data (SPMD) programming model~\cite{darema1988spmd} is a (useful) lie.
Programmers write scalar code for one logical thread and rely on the compiler~\cite{pharr2012ispc} or the hardware~\cite{lindholm2008tesla} to execute the parallel program efficiently.
But modern CPUs and GPUs are hierarchical \textit{SIMD} machines that reach peak throughput only when all lanes of a SIMD unit, each running one logical thread, execute the same instruction (\textit{instruction coherence}), access contiguous memory (\textit{memory coherence}), and perform equal amounts of work (\textit{load balance}); otherwise lanes sit idle and bandwidth is wasted.
SPMD (purposefully) hides these facts, and for regular parallelism the lie is harmless.
For \textit{divergent} programs, whose control flow, memory accesses, and unbounded iteration depend on per-thread runtime values, it is not: mapping divergent programs efficiently onto SIMD hardware is remarkably difficult~\cite{fung2007dynamicwarp, karrenberg2011wholefuncvec, laine2013wavefront, moll2018pcfl}, and the independent-threads abstraction falls apart.

Many physically-based simulation methods exhibit exactly this behavior: they are massively parallel but have ample divergence~\cite{pbrt4, sawhney2020wos, sawhney2023wost, miller2024wor}.
Monte Carlo path tracing~\cite{pbrt4}, for example, renders an image by tracing many rays through a scene and estimating the radiance reaching each pixel.
Rays that hit different materials run different shading code (instruction divergence); rays that scatter load geometry from different parts of the scene (memory divergence); and rays that bounce more times than their SIMD neighbors leave lanes idle (load imbalance).
The spatial queries finding those hits, usually accelerated by tree data structures~\cite{wald2014embree, root2026bonsai}, diverge in the same ways, as do other Monte Carlo methods~\cite{sawhney2020wos, sawhney2023wost, miller2024wor}, whose independent random walks diverge in trajectory and in the boundary functions they evaluate.
High performance on these programs requires structuring parallelism so adjacent SIMD lanes perform the same operations on adjacent data.

Known techniques for reducing divergence gain coherence at the cost of locality, parallelism, or extra work.
Wavefront path tracing~\cite{laine2013wavefront} turns the dispatch on material type into data flow through per-material work queues, making each shading stage instruction-coherent at the cost of moving live state out of registers and through global memory.
Ray reordering~\cite{meister2020reordering} recovers spatial (and therefore memory) coherence at the cost of sorting and synchronization.
Packet tracing~\cite{wald2001packet} buys memory and instruction coherence in tree traversals with extra compute.
Persistent threads~\cite{aila2009gpurt, gupta2012persistent} reduce load imbalance (given enough parallelism), at the cost of slightly more compute and global memory traffic.
%
Because the \textit{data} determines how much of each kind of divergence arises~\cite{jakob2022drjit, aila2009gpurt}, no single execution schedule is best across machines and inputs.

Hand-optimized path tracers~\cite{pbrt4, lee2017moonray, christensen2024rendermanxpu}, Monte Carlo PDE solvers~\cite{sawhney2020wos, sawhney2023wost}, and traversal libraries~\cite{wald2014embree} each compose several of these techniques with standard loop optimizations such as tiling and vectorization~\cite{bacon1994transforms, pharr2012ispc} and with explicit binding to hardware accelerators~\cite{lindholm2008tesla, nvidia_cuda_texturefetching}.
However, these systems can express that composition only by writing it into the application: the programmer no longer writes one logical thread but the SIMD-level structure of a specific machine.
Each system is thus tied to the hardware it was tuned for (NVIDIA GPUs~\cite{pbrt4}, AMD GPUs~\cite{meister2024hiprt}, x86 vector CPUs~\cite{lee2017moonray}), and exploring a different schedule means restructuring the application.

Recent programming systems make orthogonal steps in productive directions.
\textsc{Luisa}~\cite{zheng2024gpucoroutine} and Dr.Jit~\cite{jakob2022drjit} maintain an SPMD programming model but let the user instruct the compiler to split a renderer into stages for better coherence.
\textsc{Luisa} splits the program at user-annotated suspension points and lets the user pick from a fixed set of schedulers (e.g., wavefront~\cite{laine2013wavefront} or persistent threads~\cite{gupta2012persistent}), while Dr.Jit automatically decides suspension points for wavefront execution.
These are coarse-grained, global transformations, however, and neither compiler exposes other scheduling commands to compose them with.
User-schedulable languages (USLs) similarly provide data-parallel frontends, along with a scheduling language of composable loop transformations~\cite{chen2008chill, jrk2012halide, ikarashi2022exo, senanayake2020scheduling} and hardware bindings~\cite{bansal2023mosaic, ikarashi2022exo}.
%
However, USLs have primarily been developed for feed-forward array programs and lack scheduling commands targeted at divergence-mitigation.
%

Our key insight is that the hand-optimized implementations of divergent parallel programs~\cite{pbrt4, lee2017moonray, wald2014embree, sawhney2021fcpw, sawhney2026wosx, sawhney2020wos, sawhney2023wost} can be automatically generated via a directed composition of well-known program transformations from the functional programming~\cite{appel2007continuations, reynolds1972defunc} and loop optimization~\cite{bacon1994transforms, pharr2012ispc} communities.
We propose \textit{parallel dispatch graphs} (PDGs), a compilation model that represents control flow, data flow, and parallel loops in a single abstraction, enabling these transformations to compose.
PDGs mirror how graphics researchers describe their systems, as pipelines of stages connected by work queues~\cite{sugerman2009gramps, laine2013wavefront, pbrt4}, but are precise enough to serve as an intermediate representation.
A PDG models a program as a hierarchy of parallel loops, each draining either the iteration space of a programmer-written parallel for-loop or a work queue induced by scheduling transformations that turn control flow into data flow.
Loop bodies are control-flow graphs whose edges are either \textit{eager}, transferring control immediately, or \textit{deferred}, pushing a continuation onto a queue to run when that queue is drained.
%
%
%
Our model preserves the useful lie of SPMD for programmers by pairing a machine-agnostic SPMD frontend with a scheduling language that specializes the program to its hardware and data.
We make the following contributions:
\begin{itemize}
    \item Parallel dispatch graphs (PDGs), an abstraction that models divergent parallel execution as a pipeline of logical stages within a hierarchy of parallel loops. 

    \item A push-based scheduling language of composable PDG transformations: divergence mitigation~\cite{laine2013wavefront, zheng2024gpucoroutine}, standard loop transformations~\cite{bacon1994transforms}, and binding to hardware resources~\cite{ikarashi2022exo, bansal2023mosaic}.

\end{itemize}
We implement these ideas in \sys, a parallel extension of the \bonsai~\cite{root2026bonsai, gyurgyik2026decoupling} compiler for portable tree-accelerated spatial queries.
\sys uses PDGs as an intermediate representation and compiles to x86 (SSE4/AVX2/AVX-512) and NVIDIA GPU backends.
Our evaluation shows that \sys reproduces the state-of-the-art optimizations of parallel tree traversals~\cite{wald2014embree, sawhney2021fcpw}, path tracers~\cite{pbrt4, lee2017moonray}, and Monte Carlo PDE solvers~\cite{sawhney2020wos, sawhney2023wost} as distinct points in one scheduling space, matching or outperforming the hand-optimized originals (geomean speedups of $1.02$--$2.95\times$).
%
%
With \sys we find path tracing schedules that beat the expert schedules by a further $1.3\times$ on the CPU and $1.6\times$ on the GPU (geomean), up to $13\times$, and show the best schedule is often \textit{data-dependent}, motivating a scheduling language for rapid exploration of the execution space.

\textit{Scope and Limitations}.
\sys targets divergent workloads such as parallel tree traversals and parallel random walks, as opposed to \textit{irregular} workloads, where divergence comes from irregularity in the data structures, such as sparse tensor algebra~\cite{strout2018spf, kjolstad2017taco}, relational algebra~\cite{dong2026alir}, and graph computations~\cite{zhang2018graphit}.
\sys does not support irregular parallel loop nests.
Additionally, \sys does not verify user-provided parallelism annotations; a lack of independence across parallel iterations is undefined behavior.
\sys does not yet have an autoscheduler.

\section{Optimizing a Simple Path Tracer}
\label{sec:example-path-tracer}


\input{figures/lst-rtiow-render}

\noindent Consider the simple path tracer in \Cref{lst:rtiow-render}, which fills an output image \texttt{out} by sending a ray per pixel into a geometric scene and computing the illumination returned along the ray to the pixel.
A path tracer has ample parallelism, because every pixel performs independent work, and is cleanly written in SPMD style: \inputcode{render} is a parallel loop over pixels, and the loop body is scalar code for one pixel.
However, a path tracer also has ample divergence.
\inputcode{Li} first calls \inputcode{trace}, which loads nearby data from the scene data structure for rays traveling in similar directions~\cite{wald2001packet} but loses that spatial coherence after the first scattering step~\cite{aila2010treelet, pharr1997memcoherent} (memory divergence).
The \code{match} on the result of \inputcode{trace} calls \inputcode{background} for rays that miss geometry and \inputcode{interact} for rays that hit geometry, and \inputcode{interact} applies a different shading method per material (instruction divergence).
The \code{match} on the result of \inputcode{interact} recursively evaluates illumination for scattering rays, while absorbed rays return an illumination value (load imbalance).

\Cref{fig:pcg-rtiow} illustrates the parallel dispatch graphs (PDGs) of \inputcode{Li} and \inputcode{interact}: a parallel loop over pixels executes the control flow of \inputcode{Li}, which calls functions (teal boxes) and dispatches on sum types (green boxes).
The dashed arrows are the possible sources of divergence. 

\input{figures/fig-pcg-rtiow}

Consider a SIMD trace of the functions called within \inputcode{Li} across four lanes, read from left to right:
\noindent\begin{center}
\begin{tikzpicture}
    \tracerow{trace,background,,,,,,} 
    \tracerow{trace,,interact,trace,,interact,trace,background}
    \tracerow{trace,,interact,trace,background,,,}
    \tracerow{trace,,interact,trace,,interact,trace,background}
    \drawtrace
 
    \foreach \r in {0,...,\lastrow} {
        \node[anchor=east, font=\scriptsize] at (-0.1, {-(\r+0.5)*\cellheight}) {Lane \r};
    }
\end{tikzpicture}
\end{center}

\noindent Each white space, or bubble, is instruction (or occupancy) divergence, where a lane is masked out as another executes, and represents wasted work.
Optimizing this program means minimizing the bubbles, weighted by cost (e.g., thousands of cycles for a bubble alongside \inputcode{trace}).

\begin{wrapfigure}{r}{0.61\textwidth}
\vspace{-1em}
\begin{tikzpicture}
    \tracerow{trace,,,,,background} 
    \tracerow{trace,interact,trace,interact,trace,background}
    \tracerow{trace,interact,trace,,,background}
    \tracerow{trace,interact,trace,interact,trace,background}
    \drawtrace
 
    \foreach \r in {0,...,\lastrow} {
        \node[anchor=east, font=\scriptsize] at (-0.1, {-(\r+0.5)*\cellheight}) {Lane \r};
    }
\end{tikzpicture}
\vspace{-2em}
\end{wrapfigure}

A useful transformation \textit{defers} execution of \inputcode{background} until all lanes have finished tracing and shading, which results in the execution trace on the right.
For just these four lanes of work, this trace is optimal: the long-pole lanes have no bubbles.
However, the dispatch graph must be fundamentally rewired to defer execution of \inputcode{background}.
\Cref{fig:pcg-rtiow-defer-bg} shows this rewiring: the control flow that originally executed \inputcode{background} becomes \textit{data flow} into the queue \code{b}, which saves the (continuation) state needed to resume \inputcode{background}, and a second parallel loop \textit{drains} \code{b} and executes \inputcode{background}.
\Cref{lst:rtiow-render-defer-bg} shows the code \sys generates for the transformed PDG.
Note that the change is nonlocal: the value of \code{p} was not accessible within \inputcode{Li}, requiring \inputcode{render} to patch \code{p} into the queue element to complete the continuation.
The code changes considerably, but the PDG differs in one edge and the induced drain: every edge in \Cref{fig:pcg-rtiow} is \textit{eager} and transfers control immediately, while the edge into \inputcode{background} in \Cref{fig:pcg-rtiow-defer-bg} is \textit{deferred}, saving a continuation into \code{b} for the drain to execute later.

\input{figures/fig-defer-bg}

While deferral is useful in its own right, production renderers compose deferral with other transformations.
The PBRT GPU renderer~\cite{pbrt4} tiles the loop over pixels so that a single set of queues fits in device memory, defers many stages (e.g., material shaders and shadow ray queries) into work queues in the wavefront style of \citet{laine2013wavefront}, specializes the shading stage per material type, and accelerates ray tracing and texture evaluation through hardware primitives~\cite{nvidia2018turing, nvidia_cuda_texturefetching}.
\Cref{fig:pcg-rtiow-wavefront} visualizes PBRT's execution model applied to the simple renderer: each transformation is a local edit to \Cref{fig:pcg-rtiow}, and the recursion of \inputcode{Li} has become a cycle through the trace queue \code{t}.
\Cref{lst:rtiow-wavefront} shows the generated code: six parallel loops over queues and a loop that drains \code{t} until empty, which bears little resemblance to the clean SPMD specification of the algorithm.
The next section presents the conceptual model behind these transformations, showing how PDGs capture the structure of divergent parallel execution, and how scheduling commands transform a PDG.

\input{figures/fig-pcg-wavefront}

\section{Scheduling Model: Parallel Dispatch Graphs}
\label{sec:pdgs}



Parallel dispatch graphs model the \textit{structure} of divergent parallel execution: a hierarchy of parallel loops whose bodies are control-flow graphs with eager and deferred edges, where each loop drains the iteration space of a programmer-written \code{parfor} or a schedule-induced queue.
Scheduling maps this graph onto the structure of the machine.
We first describe the components of a parallel dispatch graph, then the three classes of rewrite a schedule composes.
The next section describes the programming system built on this model.

\subsection{Components of a Parallel Dispatch Graph}

A parallel dispatch graph has four transformable components:
\begin{description}
    \item[Parallel loops] drain either the range of a programmer-written \code{parfor} (\Cref{fig:pcg-rtiow} loop \code{p}) or a schedule-induced queue (\Cref{fig:pcg-rtiow-defer-bg} loop \code{b}). Nested loops form hierarchy.

    \item[Calls] transfer control to a function, whose body is a control-flow graph of its own (the teal boxes of \Cref{fig:pcg-rtiow}). A call, or its continuation, can be deferred.

    \item[Dispatches] on a sum type, including booleans, are the only conditional control flow (the \code{match} on \code{hit.material} in \Cref{fig:pcg-rtiow-interact}). A dispatch is \textit{varying} if iterations of the enclosing loop can take different arms and \textit{uniform} if all take the same arm; only varying dispatches produce SIMD divergence. Dispatch edges can also be deferred.

    \item[Edges] are \textit{eager} or \textit{deferred}. An eager edge transfers control immediately. A deferred edge is \textit{data flow}: the edge appends the live state of the iteration and a continuation to a queue, whose drain executes the continuation (the edge into \code{b} in \Cref{fig:pcg-rtiow-defer-bg}).
\end{description}
Unconditional control flow (jumps and returns) is coherent by nature and not manipulated by the schedule.
Before scheduling, a parallel dispatch graph is a control flow graph with parallel annotations, as in other parallel intermediate representations~\cite{schardl2019tapir, rainey2021parasm}; ours differs in exposing data flow as a schedulable construct.

Every parallel loop, and therefore every queue, is \textit{owned} at a level of the hierarchy, a loop or the root of the computation, and the drain of a queue is placed at the owner, after every producer.
For example, the root owns \code{b} in \Cref{fig:pcg-rtiow-defer-bg}, and the \code{band} loop owns every queue in \Cref{fig:pcg-rtiow-wavefront}.
A queue may have several producers with different continuations, in which case the stored states carry a continuation tag that the drain dispatches on.
A drain can also feed a queue upstream of itself, which happens when a schedule defers part of a recursive function; the resulting cycle executes by iterating until the queue is empty, as the \code{while !empty(t)} loop of \Cref{lst:rtiow-wavefront} does for the backedge into \code{t} in \Cref{fig:pcg-rtiow-wavefront}.
Queues that directly feed themselves are double-buffered (\Cref{sec:codegen}).

\subsection{Scheduling as PDG Rewriting}
\label{sec:rewriting-pdgs}

Every scheduling command is a \textit{transformation} of the PDG.
\sys supports three classes of transformations: loop transformations that rewrite the hierarchy, control flow transformations that mitigate divergence and specialize code, and bindings that attach execution to hardware resources.
%

\textit{Loop Transformations}.
Loop transformations create or remove hierarchy to match the hierarchy of the machine: tiling splits a loop in two, collapsing and interchange flatten and reorder nested loops, and a \textit{persistent} transformation replaces a parallel loop with a loop of constant extent whose iterations claim the original iterations from an atomic counter to load balance~\cite{aila2009gpurt, gupta2012persistent}.
These transformations apply to any parallel loop, whether the loop drains a \code{parfor} or a queue.
%

\textit{Control Flow Transformations}.
Control flow transformations mitigate divergence and specialize code.
\textit{Deferral} replaces an eager edge with a deferred edge into a queue, capturing the continuation~\cite{appel2007continuations, zheng2024gpucoroutine}; the drain then executes the continuations together, at the cost of writing the continuations to memory and reading them back.
%
%
Deferral is nonlocal:
the continuation captures the live state of every frame between the owner and the call, and each frame on that path is rewritten to report whether work was deferred, as \inputcode{render} patches \code{p} in \Cref{lst:rtiow-render-defer-bg}.
\textit{Specialization} splits a PDG (or sub-PDG) into one copy per variant of a sum type, so that each copy runs a single arm of the dispatch on that type.
On a uniform dispatch, specialization is loop unswitching~\cite{allen1972catalogue}: the dispatch moves out of the loop, which reduces register pressure and lets each arm take a different schedule.
On a varying dispatch, specialization moves the dispatch into the data flow: each iteration is routed to a per-arm queue, and each drain runs one arm alone, with no divergence at that dispatch.
\textit{Sorting} orders the iterations of a loop by a key, so that neighboring lanes take the same arms or touch nearby memory~\cite{meister2020reordering, nvidia2022ser}.
\textit{Loopification} turns a (linear or tree)-recursive function into a loop with an explicit stack~\cite{steele1977goto}.

\textit{Binding to Hardware}.
The final step of scheduling binds the PDG to the machine.
Parallel loops can be bound to a parallel resource (e.g., CPU threads or GPU blocks or threads), functions can be bound to fixed-function accelerators (RT cores~\cite{nvidia2018turing}, texture units~\cite{nvidia_cuda_texturefetching}), and queues can be assigned to a level of the machine's memory hierarchy.
Bindings must respect the hierarchy of the machine (\Cref{sec:hardware-models}), e.g., a loop bound to GPU threads must lie beneath one bound to GPU blocks.
For example, in \Cref{fig:pcg-rtiow-wavefront}, \inputcode{trace} can be assigned to the RT cores because the drain of \code{t} is bound to OptiX threads~\cite{parker2010optix}, a requirement of the NVIDIA programming model for RT cores.
Bindings are considered \textit{final}, meaning \sys rejects any further transformation of a bound loop or call.

\textit{Vectorization}.
Vectorization is a loop transformation, a control flow transformation, and a binding at once: it binds the iterations of a parallel loop to hardware vector lanes and linearizes the varying control flow within the loop~\cite{pharr2012ispc, moll2018pcfl}, preserving uniform control flow.
%

\textit{Scheduling Example}.
\Cref{lst:rtiow-render-wavefront-schedule} composes all three classes to transform \Cref{fig:pcg-rtiow} into \Cref{fig:pcg-rtiow-wavefront}, and therefore \Cref{lst:rtiow-render} into \Cref{lst:rtiow-wavefront}, in seven steps:
\begin{enumerate}
    \item Tile the \code{p} loop into an outer loop \code{band} and an inner loop \code{i}.
    \item Defer evaluation of \code{trace} into a work queue, \code{t}, owned by an iteration of \code{band}.
    \item Defer evaluation of \code{interact} into a queue also at \code{band}.
    \item Specialize the \code{interact} work queue on material type, producing work queues (\code{l}, \code{m}, \code{d}, \code{D}).
    \item Defer computation of \code{background} into a sixth work queue, \code{b}, at \code{band}.
    \item Bind the drain of \code{t} to OptiX~\cite{parker2010optix} threads, the evaluation of \code{trace} to RT Cores~\cite{nvidia2018turing}, and the evaluation of \code{eval} to the texture units~\cite{nvidia_cuda_texturefetching}.
    \item Tile and bind the \code{i} loop and drains of the remaining queues to GPU blocks and threads.
\end{enumerate}
The remaining \code{band} loop is not bound to a parallel resource, so it runs sequentially.

\vspace{0.75em}

\begin{lstlisting}[language=schedule, style=code,
%caption={Schedule for \Cref{fig:pcg-rtiow-wavefront}, as described in \Cref{sec:scheduling-example-english}.},
caption={The schedule that transforms \Cref{fig:pcg-rtiow} into \Cref{fig:pcg-rtiow-wavefront}, in the scheduling language of \Cref{sec:schedule-syntax}.},
label={lst:rtiow-render-wavefront-schedule}]
band, i = render.split(p, TILE) // split p by TILE into an outer band loop and inner i loop
t = render.queue(band, Device) // create a queue in Device memory inside of a band iteration
k0 = Li.defer(trace, t) // defer execution of trace into queue t, continuation is `k0`
q = render.queue(band, Device) // create a queue in Device memory inside of a band iteration
k1 = k0.defer(interact, q) // defer evaluation of interact into queue q
l, m, d, D = k0.specialize(q, hit.material) // q splits per sum arm of Material ADT
b = render.queue(band, Device) // create a queue in Device memory inside of a band iteration
k2 = k0.defer(background, b) // defer execution of background into queue b
render.bind(t, OptixThread) // bind the drain of t to OptixThreads
k0.bind(trace, RTCore) // bind evaluation of trace within k0 to the RT Core
interact.bind(eval, TextureUnit) // bind evaluation of eval to the texture unit
render.gpu_tile(i, 256) // syntactic sugar for x, y = split(i, 256); bind(x, GPUBlock);
                        //                     bind(y, GPUThread)
// also gpu_tile l, m, d, D, and b
\end{lstlisting}


\section{Programming System Overview}

%
We build \sys as a parallel extension to \bonsai~\cite{root2026bonsai, gyurgyik2026decoupling}, a strict, statically-typed functional query language that produces efficient tree traversals for spatial queries.
Tree traversals are themselves divergent programs~\cite{aila2009gpurt}, and are used in divergent parallel spatial Monte Carlo simulators like path tracing and spatial PDE solvers~\cite{sawhney2020wos, sawhney2023wost, miller2024wor}.
%
Thus, a compiler that abstracts away the tree data structures~\cite{root2026bonsai}, their memory layouts~\cite{gyurgyik2026decoupling}, and now their parallel execution strategies lets programmers productively explore how to optimize spatial query programs.

\Cref{fig:overview} illustrates how \sys extends the \bonsai compilation pipeline.
We inherit the set operation and algebraic data type (ADT) specification languages of \bonsai~\cite{root2026bonsai} and the Scion memory layout specification language~\cite{gyurgyik2026decoupling}.
We extend the frontend programming model with SPMD-style \textit{parallelism annotations}, whose syntax and semantics are described in \Cref{sec:parallel-annotations}.
\Cref{sec:schedule-syntax} describes the syntax of our scheduling language, in which transformations consume and produce code \textit{cursors}~\cite{ikarashi2025exo2}, pointers to code objects.

\input{figures/fig-compilation}

\Cref{sec:lowering-to-pdgs} describes the simple lowering of \sys's SPMD frontend to our PDG intermediate representation (\Cref{sec:ssa-blocks}), which runs after \bonsai has lowered set expressions to explicit tree traversals~\cite{root2026bonsai} and before it specializes memory layouts~\cite{gyurgyik2026decoupling}.
The schedule is then applied to the PDG as a series of transformations (\Cref{sec:transforms}).
Lastly, \Cref{sec:codegen} describes code generation from the transformed PDG to LLVM IR~\cite{lattner2004llvm}, including layout lowering~\cite{gyurgyik2026decoupling} and vector instruction selection.



\section{Expressing Parallel Algorithmic Logic}
\label{sec:parallel-annotations}

\input{figures/grammar-frontend}
\sys extends \bonsai with the imperative structure of SPMD programming interfaces~\cite{pharr2012ispc, lindholm2008tesla} (\Cref{fig:frontend-grammar}): standard control flow (\code{if}/\code{match}/\code{for}) and recursion, which \bonsai lowers set operations to~\cite{root2026bonsai}; a parallelism annotation, \code{parfor}; and access-restricted types.
\sys is thus a multi-paradigm language: parallel structure is exposed imperatively, while spatial queries are expressed as first-class operations over sets~\cite{root2026bonsai}.
\code{parfor} asserts that a loop's iterations are independent, expressing the \textit{possibility} of parallelism without the guarantee of parallel execution; \code{for} and \code{while} loops are sequential, and loop transformations apply only to parallel loops.

%

The programmer is responsible for the correctness of their parallel annotations: \sys does not verify that \code{parfor} iterations are independent.
Instead, inspired by Liszt~\cite{devito2011liszt}, we restrict program behavior to help the programmer write safe parallel code: all data is read-only, reduce-only, or write-only.
Aliasing is undefined behavior.
The type system enforces the access restrictions by making data immutable by default, with opt-in \code{mut} types for write-only data and \code{reduce}$(\oplus)$ types for reduce-only primitives; reductions may also be marked \code{atomic}.

\section{Schedule Language Syntax}
\label{sec:schedule-syntax}


A \sys schedule is a sequence of directives that transform a program's PDG, defining an execution order and parallelization strategy without changing \textit{what} is computed.
%
Each directive applies a transformation of \Cref{sec:rewriting-pdgs} to a \textit{cursor} (\Cref{sec:cursors}), a pointer to a piece of a PDG; some directives construct new cursors (e.g., loop splitting returns two loop cursors, given one).
%
Directives that bind to hardware are checked against a hardware model (\Cref{sec:hardware-models}).
%



\subsection{Cursors}\label{sec:cursors}
Cursors~\cite{ikarashi2025exo2} represent pointers to specific pieces of code.
In \sys, cursors are loop, function, or edge references and can come from three places: program cursors from the original program (\code{parfor} loop indices, functions, and variables within functions), schedule-induced cursors constructed by a scheduling command, and data-structure-induced cursors, where the lowering of a \bonsai~\cite{root2026bonsai, gyurgyik2026decoupling} tree traversal produces a loop (e.g., over the children of a tree node).\footnote{If tree traversals were written directly, instead of using \bonsai, these would just be program cursors.}
\Cref{lst:rtiow-render-wavefront-schedule} uses program and schedule-induced cursors, and \Cref{lst:fcpw-ch-schedule} in \Cref{sec:eval-tree-traversals} uses data-structure-induced cursors to schedule tree traversals.
\Cref{fig:schedule-syntax} includes syntax for our cursor notation, which includes loops, calls, call continuations (\schedule{g.cont}), and match arms (\schedule{x::V} for the \schedule{V} arm of the match on variable \schedule{x}).

\subsection{Transformations}\label{sec:schedule-syntax-transformations}

\Cref{fig:schedule-syntax} gives the syntax of \sys's scheduling language. \Cref{fig:schedule-signatures} gives each transformation’s signature, indicating which produce new cursors.


Loop transformations are the standard transformations seen in Halide~\cite{jrk2012halide, jrk2013halide} and Exo~\cite{ikarashi2022exo, ikarashi2025exo2}, with one exception.
\schedule{persistent(l, n)} is the persistent-thread transformation of \Cref{sec:pdgs}: \schedule{n} iterations claim the iterations of \schedule{l} from an atomic counter.
Composed with \schedule{split}, the iterations claim chunks instead, which reduces contention on the counter, as proposed by \citet{aila2009gpurt}.
Additionally, \schedule{split} has a three-loop form (omitted for brevity) that returns a third cursor for the \textit{tail loop} when the split is not exact.

\input{figures/fig-schedule-lang}

Control flow transformations are the novel scheduling language contribution of \sys.
Transforming control flow into data flow is composed of two separate transformations:
\begin{description}


    \item[Queueing] \schedule{q = f.queue(i[, M][, n])} places a queue \schedule{q}, owned at the loop \schedule{i} of \schedule{f}, in storage \schedule{M} with capacity \schedule{n}; \schedule{f.queue(root)} places the queue outside of any loop. \schedule{M} must be available at \schedule{i} in the hardware model (\Cref{sec:hardware-models}). If \schedule{n} is omitted, \sys infers it (\Cref{sec:queue-codegen}).

    \item[Deferral] \schedule{c = f.defer(e, q)} rewrites the edge \schedule{e} within \schedule{f} into an append to \schedule{q}, which must be owned within \schedule{f} or within a caller of \schedule{f}. The returned cursor \schedule{c} is the continuation the drain of \schedule{q} executes.

\end{description}
If several \schedule{defer} directives target one queue, the drain dispatches to their continuations (\Cref{sec:transforms}).


Other control flow transformations include:
\schedule{specialize(c, x)}, which splits the cursor \schedule{c} per variant of the ADT \schedule{x} and returns one cursor per variant (e.g., the per-material queues in \Cref{lst:rtiow-render-wavefront-schedule}), and can produce unswitched loops for uniform control flow (\Cref{sec:transforms}).
\schedule{sort(c, k)} orders the instances of \schedule{c} by the key \schedule{k}: e.g., the rays in a queue by a spatial hint before tracing or the recursive calls of a tree traversal (\Cref{sec:eval-tree-traversals}).
Lastly, \schedule{loopify([n][, M])} turns a tail-recursive function into a loop~\cite{steele1977goto}; for tree recursion, \schedule{n} and \schedule{M} size and place a software stack (\Cref{sec:eval-tree-traversals}).


Finally, \sys supports hardware binding in the style of Exo~\cite{ikarashi2022exo} and Mosaic~\cite{bansal2023mosaic}.
\schedule{bind(c, R)} executes the cursor \schedule{c} on the resource \schedule{R}: a loop on CPU threads or GPU blocks or threads, or a function on an accelerator such as the RT cores~\cite{nvidia2018turing} or texture units~\cite{nvidia_cuda_texturefetching}.
\schedule{prefetch(l, n)} prefetches \schedule{n} cache lines ahead of the loop \schedule{l}, as in Halide~\cite{jrk2012halide}.
\schedule{gpu_tile(l, n)} is syntactic sugar for \schedule{x, y = split(l, n); bind(x, GPUBlock); bind(y, GPUThread)}.

\input{figures/fig-hardware-models}

\subsection{Hardware Models}\label{sec:hardware-models}


A \textit{hardware model} describes a machine as a hierarchy of compute resources $R$ and the storage kinds $M$ each level owns.
\schedule{bind}, \schedule{queue}, and \schedule{loopify} are checked against the model in lowering: storage of kind $M$ is owned at a level that lists $M$ and used only at or beneath its owner.
For example, an \schedule{RTCore} binding must lie within an \schedule{OptixThread} loop~\cite{parker2010optix, nvidia2018turing}, and a \schedule{DeviceShared} queue within the block that owns it.
\schedule{Device} is the exception: the host owns it, as \schedule{band} owns the queues of \Cref{lst:rtiow-render-wavefront-schedule}, but only GPU loops use it.
The design follows Sequoia~\cite{fatahalian2006sequoia} and Exo~\cite{ikarashi2022exo}, and can easily extend to other machines.

\section{Parallel Dispatch Graph Intermediate Representation}
\label{sec:ir}

We present PDGs as an SSA intermediate representation (IR), outline the lowering from \sys's frontend to PDGs, and show how schedule transformations are expressed as IR rewrites.



\subsection{SSA Blocks with Terminators}
\label{sec:ssa-blocks}

We implement PDGs as an SSA intermediate representation with explicit block arguments~\cite{appel1998ssa}.
Block terminators represent all control flow, parallelism, and data flow, and explicitly pass continuation parameters.
%
The live state at a call is therefore the argument list of the continuation block, and continuation capture needs no liveness analysis~\cite{kildall1973dataflow}.
Most transformations (other than vectorization) solely manipulate the terminators of blocks, as \Cref{sec:transforms} will show.
Instructions within a block are arithmetic, side-effecting operations such as stores, and explicit vector operations, as in LLVM~\cite{lattner2004llvm} and Halide~\cite{jrk2012halide, jrk2013halide}.
%
%
\sys's IR has 8 terminators:
\begin{description}
    \item[Jump] \irlang|jmp| terminators are standard unconditional jumps. They pass arguments from the source block's scope to the target block.

    \item[Conditional Jump] \irlang|cjmp| terminators are n-ary conditional jumps on a sum type (the dispatches of \Cref{sec:pdgs}) and the only conditional control flow in the IR. The variables bound by each arm are prepended to the arguments of that arm's block.

    \item[Call] \irlang|call| terminators transfer control flow to a function target, also expressing the return continuation block. If the function has a return type, the return value is prepended to the continuation's argument list.

    \item[Return] \irlang|ret| terminators return control flow to the caller, optionally returning a value.

    \item[Multicall] \irlang|mcall| terminators are a sequence of calls to a function, representing \bonsai's \code{from} node~\cite{root2026bonsai}: unordered tree recursion on a static set of arguments. This node simplifies some pattern matching in \Cref{sec:sort-transform}.

    \item[Parallel] \irlang|par| terminators take an index name, an iteration slice (start/end/stride), a loop body block (which takes the loop index as the first argument), and a continuation block. The arguments to the loop body block (other than the index) are \textit{uniform} arguments.


    \item[Yield] \irlang|yield| terminators end parallel loop bodies and take no arguments.

    \item[Defer] \irlang|defer| saves a continuation to a named queue and then yields control flow, returning the queue index if the defer is nested within a callee.
\end{description}

\begin{wrapfigure}{r}{0.6\textwidth}
\vspace{-2em}
\begin{lstlisting}[language=terminators, mathescape=true, basicstyle=\scriptsize\ttfamily, backgroundcolor=\color{codebackground}, frame=single, linewidth=\linewidth, caption={PDG block terminators; optional fields are marked \code{?}.}, label={lst:terminators}]
jmp block(args)
cjmp v [block0(args0), ..., block{n-1}(args{n-1})]
call f(argsf) cont(argsb) hw_tag?
ret val?
mcall f [args0, ..., args{n-1}] mask? order? cont(argsb)
par idx range body(argsu) cont(argsc) hw_tag?
yield
defer args queue
\end{lstlisting}
\vspace{-1.5em}
\end{wrapfigure}

\noindent \Cref{lst:terminators} summarizes these terminators.
%
In listings, a continuation written as \irlang|ret| or \irlang|yield| is an empty block ending in that terminator.
%
Note that any results of parallelism must be communicated via explicit memory stores and atomics: \irlang|par| does not return a value.

\subsection{Lowering to PDGs}
\label{sec:lowering-to-pdgs}

Lowering an input program into a PDG is straightforward, via the SSA construction of \citet{braun2013ssa}.
\code{match} lowers to \irlang|cjmp|, as does any conditional control flow (\code{if} is a \irlang|cjmp| on the condition); calls and returns lower to \irlang|call| and \irlang|ret|.
\code{parfor} lowers to \irlang|par|, with the body of the loop as the body block and the code after the loop as the continuation.
\bonsai's \bonsaicode{from} lowers to \irlang|mcall|.
\Cref{lst:rtiow-ir} in Appendix~\ref{appendix:rtiow-ir} shows the lowering of \Cref{lst:rtiow-render} after simplification.


\subsection{Transforming PDGs}
\label{sec:transforms}

\textit{Loop Transformations}.
%
Loop transformations other than vectorization rewrite the \irlang|par| terminator the cursor points to and leave the loop body alone~\cite{bacon1994transforms, jrk2012halide}.
Vectorization also rewrites the body: we implement Moll and Hack's partial control-flow linearization~\cite{moll2018pcfl}, which provably preserves uniform control flow where possible.

\textit{Queue}.
\schedule{queue} places a symbolic allocation at the cursor.
The queue records every \schedule{defer} that pushes into the queue.
The \schedule{defer} transformation places the drain and \Cref{sec:queue-codegen} sizes the allocation.


\textit{Defer}.
Deferral, \schedule{c = f.defer(e, q)}, is a selective CPS conversion~\cite{appel2007continuations} of the call path from the loop that owns \code{q} down to \code{e}, followed by defunctionalization~\cite{reynolds1972defunc} of the continuation into the state stored in \code{q}.
Instead of executing the edge \code{e}, the producer stores the live state of every frame on the path into \code{q} and ends execution.
The drain of \code{q} loads that state and executes the deferred control flow.
Every function on the path to the deferred edge either returns its value (\code{Done}) and executes normally, or the index of the stored state (\code{Defer}) and saves the live state within the frame at that index.
In \Cref{lst:rtiow-render-defer-bg}, the path runs from the loop over \code{p} through \code{Li} to \code{background}: \code{Li} stores \code{L}, \code{beta}, and \code{ray} in \code{b} and returns the index, \code{render} adds \code{p}, and the drain of \code{b} finishes both functions.
In the IR, \code{e} becomes a \irlang|defer| of its arguments and its continuation's arguments, and the body of the owning loop yields.
The live state of a frame is the arguments of its continuation block, except cheap (and pure) recomputable values~\cite{briggs1992remat}.
Tail calls leave no live state, e.g., the recursive call in \code{Li}.
Non-tail recursive calls on the path require unbounded state; \sys rejects defer commands that cannot be statically bounded.
The drain is placed at the owner, after all producers, and is a parallel loop over \code{q} that executes \code{e} and the remainder of the frames of the continuation; this drain code is the cursor \code{c}.
A queue with several producers stores a continuation tag with each state, on which the drain dispatches; a function that defers on every path drops \code{Done} and its dispatch, as in \Cref{lst:rtiow-wavefront}.
\Cref{alg:defer} in Appendix~\ref{appendix:defer-impl} gives the full algorithm for this transformation.

%
\sys's type restrictions (\Cref{sec:parallel-annotations}) make legality simple to check: deferral is legal as long as no statement between a producer of a queue and its drain changes the access type of data the deferred continuations touch (e.g., reads data the deferred edge writes).
Within a \code{parfor}, iterations are independent and each datum is read-only, write-only, or reduce-only, so pieces of its iterations may run in any order (up to order of reductions); only a statement after the loop can read what an iteration of the loop wrote.
\sys therefore checks every statement that can run between a producer and the drain against the data the continuation accesses: each must maintain the access type it has in the producer, otherwise the transformation is rejected.

\textit{Specialize}.
%
Specialization takes a cursor and an ADT variable in scope at that cursor, and returns one copy of the cursor per variant.
\schedule{l, m, d, D = k0.specialize(q, hit.material)} in \Cref{lst:rtiow-render-wavefront-schedule} splits the appends into \code{q} by material and the drain of \code{q} into per-material drains, a specializing defunctionalization~\cite{brandon2023defunc, reynolds1972defunc}.
For a loop and a uniform variable, specialization is loop unswitching~\cite{allen1972catalogue}.
In our implementation of PBRT's renderer, for example, the parallel loop over pixels dispatches on the integrator that computes radiance; specializing on the integrator ADT yields one parallel loop per integrator, each with a different schedule.

\textit{Sort}.
\label{sec:sort-transform}
Sorting has different lowerings for different cursors.
Sorting a loop over a queue lowers to vendor-provided sort implementations~\cite{tbbsort, cccl}.
On a function within a loop body, sort reorders the iterations of the loop body, e.g., via OptiX's shader execution reordering~\cite{nvidia2022ser}.
On an \irlang|mcall|, recursion into a tree's children~\cite{gyurgyik2026decoupling}, the sort fixes the order in which the children are visited, which is always legal for traversals lowered by \bonsai~\cite{root2026bonsai}, and lowers to a sorting network~\cite{adams2021median}.


\textit{Loopify}.
Loopify applies to a tail-recursive function, linear or tree.
For linear recursion, the standard tail-call optimization~\cite{steele1977goto} rewrites \irlang|call f(args) ret| into \irlang|jmp entry(args)|.
For tree recursion, the recursion \bonsai produces~\cite{root2026bonsai}, loopification places an explicit stack of the given size at the entry of the function and rewrites \irlang|mcall f [...] ret| into pushes onto the stack followed by a jump to the loop head.



\textit{Bindings}.
Bindings attach hardware tags to function calls and loop bodies in the PDG.
\sys reports an error if the target is already tagged or if a further transformation is applied to the cursor.


\section{Code Generation and Optimization}
\label{sec:codegen}


\sys applies standard optimizations~\cite{wegman1991cprop,knoop1992lcm,mueller1995coderep,knoop1994pdce} before lowering to LLVM IR~\cite{lattner2004llvm} by translating block arguments to $\phi$-nodes.
We describe the non-standard parts of code generation below.


\textit{Layouts}.
The final lowering to LLVM also stamps in concrete memory layouts for tree data structures and software queues.
For tree data structures, \sys uses the same lowering machinery as ~\citet{gyurgyik2026decoupling}, which supports user-specified memory layouts for ADT-defined accelerator trees~\cite{root2026bonsai}.
For queues, \sys generates a struct-of-arrays (SoA), which enables vectorized/coalesced loads and stores to the queue, as noted in multiple wavefront renderers~\cite{laine2013wavefront, lee2017moonray, pbrt4}.
We do not implement the more complex queue layouts described by the MoonRay renderer~\cite{lee2017moonray}, and instead match the layouts of the renderer we compare to in evaluation; the PBRT wavefront renderer uses SoA~\cite{pbrt4}.
We leave exploration of memory layouts for the work queues, which could be co-designed with the schedule, as future work.

\textit{Vectorization}.
%
%
Vector instruction selection incorporates a number of peephole optimizations inspired by ispc~\cite{pharr2012ispc} and Halide~\cite{jrk2012halide, jrk2013halide}.
%
%
For example, we use bounds analysis to lower integer division to floating point division when it is exact.
We additionally support several optimizations for vectorization over tree data structures expressed in the layout language (e.g., transforming gathers on contiguous layouts into vector loads).

\textit{Queues: Appends, Size Inference, Double Buffering, and Allocation Hoisting}.
\label{sec:queue-codegen}
An append is an atomic add on the queue's counter that reserves an index, followed by a store of the state at that index.
The drain iterates over $[0, \mathit{count})$ and resets the counter.
\sys computes an upper bound on the size of a queue as the sum of the loop extents that can produce work for the queue.
For single-producer queues (the common case in rendering workloads), this inference is simply the size of the single producer loop.
For more complex patterns, e.g., multiple producers feeding the backedge to \code{t} in \Cref{fig:pcg-rtiow-wavefront}, dataflow analysis is used to upper bound the size (e.g., \code{t} cannot grow larger than its previous size, as there is at most one unique path to insert into \code{t} from the drain of \code{t}).
A queue that appends directly to itself (e.g., through a recursive call to the deferred function) is double-buffered, with a read-only and a write-only copy at every step.
Queue allocations are hoisted when possible: the queues in \Cref{lst:rtiow-wavefront} are allocated once outside the sequential loop over bands and reused across bands.

\textit{Hardware Binding}.
As in Halide, \sys launches parallel kernels via static closure capturing.
Additionally, \sys targets NVIDIA's RT Cores~\cite{nvidia2018turing} and OptiX~\cite{parker2010optix} via the same code generation approach as Dr.Jit~\cite{jakob2022drjit}.
Texture units are targeted via the API provided by NVIDIA~\cite{nvidia_cuda_texturefetching}.
Extending \sys to support additional hardware models requires extending the hardware binding support via techniques similar to Mosaic~\cite{bansal2023mosaic} and Exo~\cite{ikarashi2022exo}.


\section{Evaluation}





%
The primary goal of our evaluation is to show that \sys's programming model offers expressive power \textit{without} sacrificing performance or portability.
We show that schedules from hand-optimized parallel tree traversals~\cite{wald2014embree, sawhney2021fcpw}, path tracers~\cite{pbrt4}, and Monte Carlo PDE solvers~\cite{sawhney2020wos, sawhney2023wost, sawhney2026wosx} can be concisely expressed in our programming model.
Furthermore, we show that the performance of our generated code matches or exceeds hand-optimized CPU and GPU implementations.
Lastly, we show that the optimal choice of schedule is \textit{data-dependent}, which strengthens the justification for providing a scheduling language for productive design space exploration of the scheduling space.

\subsection{Methodology}

\textit{Hardware and software}.
All benchmarks were measured on an AMD Ryzen~9~9950X3D with 16 Zen~5 cores (32 threads with SMT), 48\,KiB of L1 data cache and 1\,MiB of L2 per core, 128\,MiB of L3 across its two dies, and AVX-512/AVX2/SSE4 support, with 192\,GB of DDR5, and an NVIDIA GeForce RTX~5090 (32\,GB of GDDR7).
The machine runs Ubuntu~24.04.4, and uses NVIDIA driver~615.71.09 (CUDA~13.4), CUDA toolkit~13.2 and OptiX~9.1.
\sys compiles with LLVM~23.1.1, all generated code is compiled at \texttt{-O3} for the native target, and generated GPU code goes through LLVM's NVPTX backend.
All benchmark drivers are compiled with Clang~22.1 at \texttt{-O3 -march=native -std=c++20}.
%
We compile reference systems with their release configurations (\texttt{-O3 -march=native}): PBRT-v4~\cite{pbrt4}, Embree~4.4.1~\cite{wald2014embree}, FCPW~\cite{sawhney2021fcpw}, and WoSX~\cite{sawhney2026wosx}.\footnote{\url{https://github.com/mmp/pbrt-v4}, commit \texttt{eef3a6e}, \url{https://github.com/RenderKit/embree}, tag \texttt{v4.4.1}, \url{https://github.com/rohan-sawhney/fcpw}, commit \texttt{3b69ddb}, v1.2.0, \url{https://github.com/nv-tlabs/wosx}, commit \texttt{2716c27}}
Fast math is disabled everywhere except for PBRT's GPU kernels, which PBRT builds with \texttt{-{}-use\_fast\_math}; we match it on that comparison.

We match the reference system's tree exactly (topology and memory layout) to isolate the effects of the schedule for each benchmark.
Section~\ref{sec:conclusion} discusses the interplay of layout and schedule.

\textit{Timing}.
Each measurement is the minimum of three runs after one warm-up.
Times for all systems exclude loading the input and tree construction.

\textit{Verification}.
Every result is checked against the reference system's: traversal and solver outputs directly, rendered images by per-pixel difference.

\textit{Datasets}.
We evaluate on scenes and meshes from \texttt{pbrt-v4-scenes}\footnote{\url{https://github.com/mmp/pbrt-v4-scenes}, commit \texttt{30cf4a0}}, plus the six demo meshes of the WoSX repository~\cite{sawhney2026wosx} for the PDE solvers; all are summarized in Appendix~\ref{sec:appendix-statistics}.

\textit{AI Use}.
We used Claude Fable 5.1~\cite{anthropic2026fable51} for code development of some program transformations, schedule sweeps, and translation of the PBRT renderer into \sys's frontend.
All code was reviewed by the authors.
All results are verified to match up to floating point accumulation error.

\begin{figure}[t]
  \centering
  \begin{subfigure}[b]{\textwidth}
      \centering
      \includegraphics[width=\textwidth]{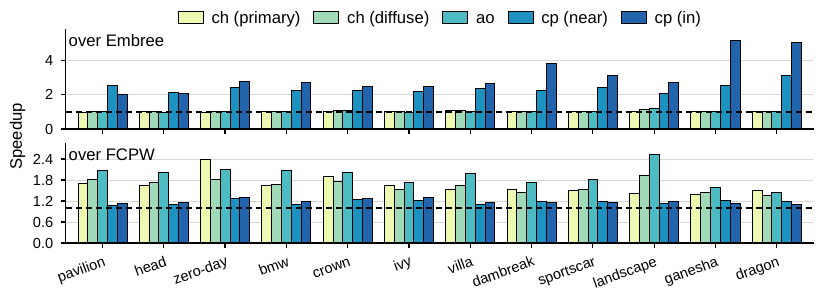}
      \caption{CPU (1 core): Embree (8-wide BVH) and FCPW (4-wide). Geomean speedup over Embree: ch 1.02$\times$/1.05$\times$ (primary/diffuse); ao 1.04$\times$; cp 2.36$\times$/2.95$\times$ (near/in). Over FCPW: ch 1.62$\times$/1.63$\times$; ao 1.90$\times$; cp 1.17$\times$/1.19$\times$.}
      \label{fig:cpu-queries}
  \end{subfigure}
  \vfill
  \begin{subfigure}[b]{\textwidth}
      \centering
      \includegraphics[width=\textwidth]{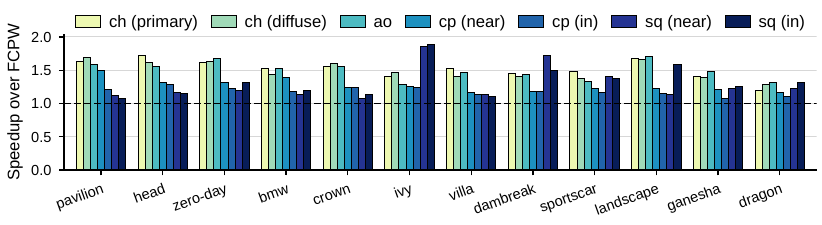}
      \caption{GPU: FCPW (binary BVH). Geomean speedup: ch 1.53$\times$/1.49$\times$ (primary/diffuse); ao 1.49$\times$; cp 1.25$\times$/1.18$\times$ (near/in); sq 1.27$\times$/1.30$\times$ (near/in).}
      \label{fig:fcpw-gpu}
  \end{subfigure}
  \caption{Speedup of \sys-generated traversals over hand-optimized libraries with matching trees and schedules (higher is better, dashed line is parity). Queries: closest hit (ch) over primary and diffuse rays, any hit (ao) queries over shadow rays, and closest point (cp) and silhouette (sq), on points near or inside the mesh.}
  \label{fig:rtq}
\end{figure}

\subsection{High-Performance Tree Traversals}
\label{sec:eval-tree-traversals}

We compare \sys-generated traversals against two hand-optimized libraries: Embree~\cite{wald2014embree} and Fastest Closest Points in the West (FCPW)~\cite{sawhney2021fcpw}.
We evaluate closest-hit (ch), any-hit (ao; ambient occlusion), closest-point (cp), and (GPU/FCPW only, not supported by Embree) silhouette (sq) queries on 12 meshes, with primary and diffuse rays for ray queries and points near or inside the mesh for point queries.
Following the benchmarking of \citet{aila2009gpurt}, each mesh is traced from a $1024\times1024$ image of primary rays, and every primary hit casts one ambient-occlusion ray and one diffuse ray.
Point queries (cp and sq) use points scattered uniformly over the bounding box, as FCPW's own test harness does~\cite{sawhney2021fcpw}, and points near the surface, as in the sampling scheme of~\citet{sawhney2020wos}.
\Cref{fig:rtq} reports speedups over each library with matching schedules (\Cref{lst:fcpw-ch-schedule}), so the differences are due to code generation alone.

\textit{CPU}.
\Cref{fig:cpu-queries} compares against the fastest CPU tree each library provides: Embree's 8-wide and FCPW's 4-wide BVH, both with hand-vectorized traversals.
Both libraries leave parallelism across queries to the application, so we measure one core to compare the quality of the vectorized traversals.
%
On the queries each library was built for, \sys matches Embree on ray queries (ch and ao; worst case 0.95$\times$, ao on ivy) and is modestly faster than FCPW on closest point queries.
On the rest, \sys is faster on every mesh: up to 5.12$\times$ over Embree on cp (in) on ganesha, and up to 2.51$\times$ over FCPW on ao on landscape.
We attribute the speedup over Embree to better vectorization of the loop over leaf data (\schedule{triangles::Leaf.data}) and the speedup over FCPW to better code generated for the vectorized sort of children (\schedule{trace.sort}).
\Cref{lst:fcpw-ch-schedule} shows the ch schedule matching FCPW’s 4-wide traversal.
Embree’s is similar, but it vectorizes over 8 children, prefetches in the loop over \schedule{triangles::Interior.cdrn}, and uses a larger stack depth.

\begin{lstlisting}[language=schedule, mathescape=true, basicstyle=\scriptsize\ttfamily, backgroundcolor=\color{codebackground}, frame=single, caption={Schedule matching FCPW's 4-wide closest-hit traversal, used for \Cref{fig:cpu-queries}.}, label={lst:fcpw-ch-schedule}, xleftmargin=3.4pt, xrightmargin=3.4pt]
trace.sort(trace, |i:u32, r:Ray, bs:vec[AABB, 4]| dist(r, bs[i])) // visit in distance order
trace.vectorize(triangles::Interior.cdrn)   // vectorize child intersection and sort
trace.vectorize(triangles::Leaf.data)       // vectorize the loop over leaf data
trace.loopify(96)                           // recursion -> loop; stack depth 96 (FCPW's max)
\end{lstlisting}


\textit{GPU}.
FCPW's GPU backend supports only a binary BVH.
We compare to it in \Cref{fig:fcpw-gpu}.
\sys is never slower than FCPW (minimum of $1.08\times$ speedup across all queries).
Geometric means of the speedups range from $1.18\times$ on closest-point (in) to $1.53\times$ on closest-hit (primary).
Profiling attributes the speedup to \schedule{loopify}: the loop \sys generates caches the top of the traversal stack~\cite{ertl1995stackcache}, reducing memory traffic relative to FCPW's hand-written loop.

\subsection{Physically Based Rendering}
\label{sec:eval-pbrt}

\begin{figure}[b]
    \centering
    \includegraphics[width=\textwidth]{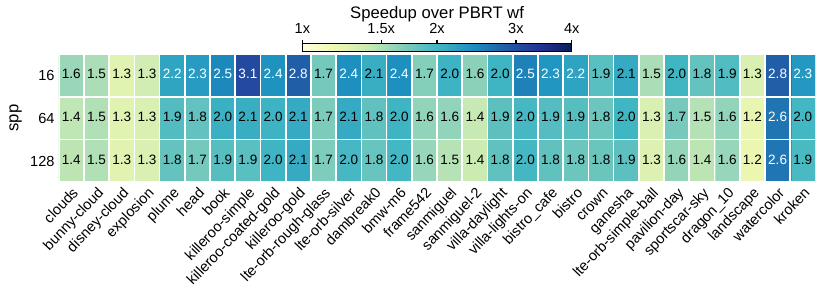}
    \caption{Speedup of \sys-generated code over PBRT's GPU wavefront with a matching schedule.}
    \label{fig:pbrt-gpu-heatmap}
\end{figure}

PBRT is a research rendering system based on path tracing, an algorithm based on Monte Carlo simulation of a ray optics model of light~\cite{pbrt4}.
This approach is the current state-of-the-art in computer graphics for image synthesis.
We compare against PBRT~\cite{pbrt4}'s volumetric path tracer, which was implemented on both the GPU and the CPU, to provide evidence for three claims:
\begin{enumerate}
    \item \sys can express complex programs and complex schedules.

    \item \sys can outperform a hand-optimized renderer.

    \item The best choice of schedule is \textit{scene-dependent}, in addition to machine-dependent.
\end{enumerate}
\Cref{fig:pcg-pbrt-wavefront} in Appendix~\ref{sec:appendix-pbrt-pdg} illustrates the parallel dispatch graph of PBRT's GPU wavefront renderer, which executes a wavefront per pixel tile of the image.
\sys's implementation of this PDG is 64 scheduling commands, where the GPU backend of PBRT took several months of development and approximately four thousand lines of code. Further, restructuring a significant amount of the C++ application code was required to run on both the CPU and the GPU~\cite{pbrt4}.
PBRT is the largest program we evaluate, and while most transformations are whole-program transformations (and the \sys compiler itself is not parallelized), compiling the PBRT schedules takes only minutes, which is comparable to PBRT's sequential compile time (\Cref{fig:pbrt-compile-times} in Appendix~\ref{appendix:compile-times}).

\begin{figure}[t]
  \centering
  \begin{subfigure}[b]{\textwidth}
      \centering
      \includegraphics[width=\textwidth]{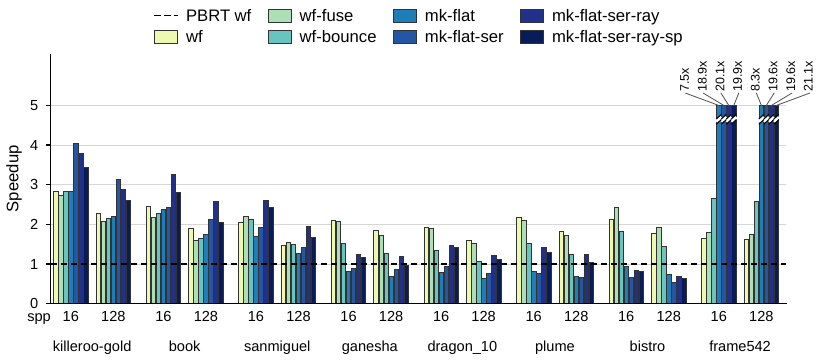}
      \caption{GPU schedule exploration: speedup over PBRT's wavefront. \texttt{wf} is \sys's matching schedule to PBRT. The best schedule is $1.56\times$ faster than \sys's \texttt{wf} (geomean).}
      \label{fig:gpu-pbrt-sweep}
  \end{subfigure}
  \vfill
  \begin{subfigure}[b]{\textwidth}
      \centering
      \includegraphics[width=\textwidth]{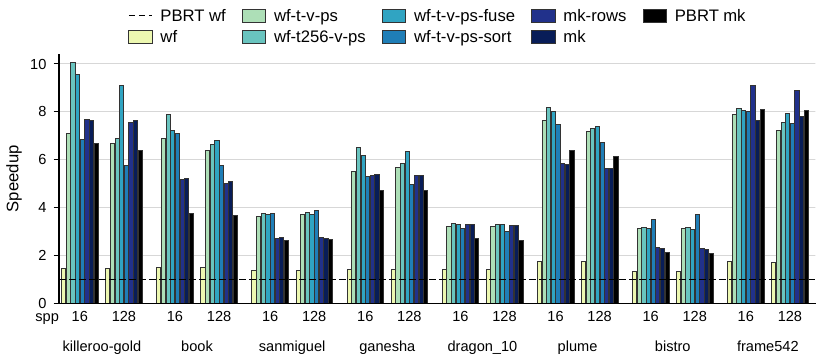}
      \caption{CPU schedule exploration (32 threads): speedup over PBRT's wavefront. \texttt{wf} and \texttt{mk} match PBRT's wavefront and megakernel (black bar). The best schedule is $1.42\times$ faster than PBRT's megakernel, $5.89\times$ faster than PBRT's wavefront, and $3.93\times$ faster than \sys's matching wavefront (geomeans).}
      \label{fig:cpu-pbrt-sweep}
  \end{subfigure}
    \caption{Schedule exploration across two hardware platforms. \texttt{wf} and \texttt{mk} match PBRT's wavefront and megakernel. The other schedules vary tiling (\texttt{-t}, \texttt{-t256}, \texttt{-rows}, and \texttt{-flat}), loop order (\texttt{-ps}, \texttt{-sp}), vectorization (\texttt{-v}), deferral (\texttt{-fuse}, \texttt{-bounce}), and sorting (\texttt{-sort}, \texttt{-ser}, and \texttt{-ray}).}
  \label{fig:pbrt}
\end{figure}

\textit{PBRT GPU Wavefront Comparison}.
\Cref{fig:pbrt-gpu-heatmap} shows the speedup of \sys-generated code over PBRT's matching renderer, across 30 scenes and 3 different samples per pixel parameters (spp roughly corresponds to work-per-pixel, which also improves image quality).
\sys is faster across all configurations.
Profiling reveals the speedup to be a memory bandwidth win: \sys saves smaller continuations than the PBRT authors manually wrote, so queues are smaller and less data is loaded and stored across each otherwise-equal rendering stage (\Cref{tab:queue-bytes} in Appendix~\ref{appendix:queue-size}).

\textit{PBRT Schedule Exploration}.
We now turn to exploration of schedules for PBRT's renderer.
\Cref{fig:gpu-pbrt-sweep} shows the speedup of 7 different schedules over PBRT's wavefront-only GPU renderer on 8 scenes chosen to span the dataset in geometry, materials, lights, and path depth (\Cref{tab:dataset-statistics}), at two samples-per-pixel (spp) rates.
All schedules (including PBRT's) map the ray trace call to the GPU's RT Cores, and \texttt{wf} is our match of PBRT's schedule (\Cref{fig:pbrt-gpu-heatmap}).
The schedules we explored include removal of different deferrals (e.g., \texttt{wf-fuse} fuses material evaluation into the trace kernel), and various megakernel (totally fused) variations, with and without shader execution reordering (SER)~\cite{nvidia2022ser}.
The PBRT-inspired schedule is not the strict winner: \texttt{wf} is the best schedule on three scenes, an SER megakernel on four, and \texttt{wf-fuse} on \texttt{bistro}.
The best schedule is $1.56\times$ faster than \texttt{wf} by geomean.
On \texttt{frame542}, the SER megakernels are $12\times$ and $13\times$ faster than \texttt{wf} at 16 and 128 spp ($20\times$ and $21\times$ faster than PBRT).
Profiling shows that a very small number of incredibly long running paths per pixel tile leaves every wavefront tile long-running but low-occupancy; the megakernel launches all paths together, and SER compacts the long-running threads.

\Cref{fig:cpu-pbrt-sweep} shows the speedup of 7 CPU schedules versus PBRT's CPU wavefront renderer on the same 8 scenes and sample rates.
PBRT also offers a megakernel renderer on the CPU (black bar), which significantly outperforms PBRT's CPU wavefront.
Our matching schedules (\texttt{wf} and \texttt{mk}) reproduce both, and \texttt{mk} is $1.10\times$ faster than PBRT's megakernel by geomean.
The best schedule is $1.42\times$ faster than PBRT's megakernel by geomean ($1.10$--$2.10\times$) and $1.29\times$ faster than \texttt{mk}.
On seven of the eight scenes the best schedule is a per-thread vectorized wavefront inspired by the MoonRay renderer~\cite{lee2017moonray}, which tiles the image into small tiles such that work queues fit into L2 cache, executes tiles in parallel on CPU cores, and uses queues to group work into vector lanes worth of work.
The fastest variant depends on the scene and sample rate: small tiles (\texttt{wf-t256-v-ps}) at 16 spp, fused material evaluation (\texttt{wf-t-v-ps-fuse}) at 128 spp, and sorting (\texttt{wf-t-v-ps-sort}) on the two scenes with the most materials.
%





\begin{figure}[t]
\centering
\begin{subfigure}[b]{\textwidth}
    \centering
    \includegraphics[width=\textwidth]{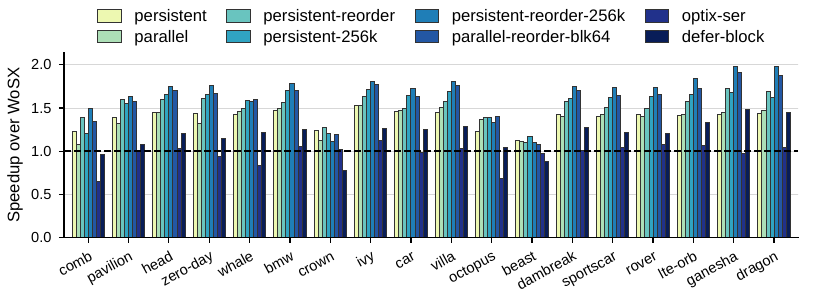}
    \caption{Walk on spheres~\cite{sawhney2020wos}. Geomean speedup: (matching) 1.38$\times$; (best) 1.64$\times$ (up to 1.98$\times$).}
    \label{fig:gpu-wos}
\end{subfigure}
\vfill
\begin{subfigure}[b]{\textwidth}
    \centering
    \includegraphics[width=\textwidth]{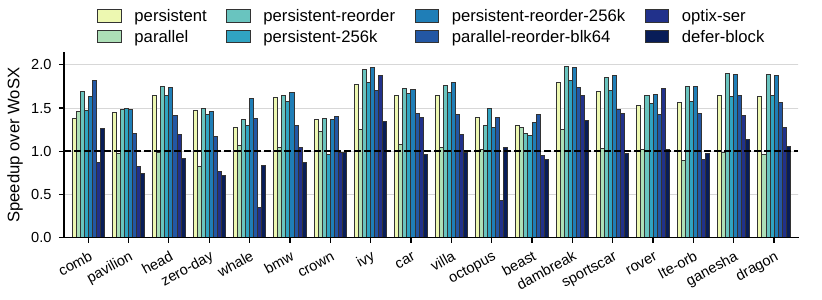}
    \caption{Walk on stars~\cite{sawhney2023wost}. Geomean speedup: (matching) 1.54$\times$; (best) 1.66$\times$ (up to 1.98$\times$).}
    \label{fig:gpu-wost}
\end{subfigure}
\caption{Speedup of \sys-generated GPU solvers over WoSX~\cite{sawhney2026wosx} on 18 inputs (higher is better, dashed line is parity). \texttt{persistent} matches WoSX's kernel; other schedules make changes explained in
Appendix~\ref{sec:appendix-wosx-schedule}.}
\label{fig:wosx}
\end{figure}

\subsection{Monte Carlo PDE Solving}
\label{sec:eval-wosx}


Walk on spheres (WoS)~\cite{sawhney2020wos} and walk on stars (WoSt)~\cite{sawhney2023wost} are Monte Carlo PDE solvers that, like a path tracer, estimate the solution at each point by averaging many independent random walks, each of which takes a data-dependent number of steps that perform spatial queries and evaluate the PDE's boundary functions.
We compare against WoSX~\cite{sawhney2026wosx}, the GPU implementation of both solvers.
Unlike PBRT's wavefront GPU renderer, WoSX is a persistent-thread kernel~\cite{aila2009gpurt}: it launches enough threads to fill the machine, which each claim a walk from an atomic counter in an effort to load balance walks across threads.
This schedule is four scheduling commands in \sys (\Cref{lst:wosx-schedule}).

\textit{WoSX Comparison}.
\Cref{fig:wosx} shows the speedup of \sys-generated code over WoSX across 18 meshes.
With the matching schedule (\texttt{persistent}), \sys is faster on every mesh: $1.38\times$ by geomean on WoS and $1.54\times$ on WoSt.
Profiling attributes the win to occupancy.
WoSX compiles a single kernel that supports every run-time option of its solver (e.g., also estimating gradients); on these benchmarks 70\% of its instructions never execute, yet the GPU compiler still reserves registers for them, so WoSX needs 82 registers per thread.
\sys's kernel, specialized to the program, needs 54 (the code WoSX actually executes needs the same), which raises achieved occupancy from 32\% to 47\% and hides more memory latency.

\textit{WoSX Schedule Exploration}.
%
The other schedules in \Cref{fig:wosx} vary the schedule (Appendix~\ref{sec:appendix-wosx-schedule}), and as with rendering, the best schedule is data-dependent.
On WoS, persistent threads make little difference: a plain launch of one thread per walk (\texttt{parallel}) is within 1\% of \texttt{persistent}.
The best schedule, \texttt{persistent-reorder-256k}, instead has each warp claim the walks of one point, which start from the same position, and launches twice as many threads; it is fastest on 14 of 18 meshes and $1.64\times$ faster than WoSX by geomean.
On WoSt, whose walks are longer and more varied in length, persistent threads make a large difference (\texttt{parallel} is $0.69\times$ as fast as \texttt{persistent}), and five different schedules are the fastest on at least one mesh.

We also explored wavefront schedules that defer each step of a walk into a queue; \Cref{fig:wosx} shows only the best, \texttt{defer-block}, which is $0.85\times$ as fast as \texttt{persistent} on WoS and $0.64\times$ on WoSt.
Wavefront execution performs poorly on this benchmark because the boundary functions of our test PDE are simple: a step is a few tree traversals, which diverge no matter which walks share a warp, and little else, so the divergence a deferral removes is not expensive enough to warrant the memory traffic of saving and restoring the step.
However, a PDE with complex (divergent) boundary functions could justify wavefront execution, as materials do for PBRT.

\section{Related Work}












\textit{Divergence mitigation in renderers and traversals}.
Many of the transformations \sys exposes as scheduling directives were first described in manual restructurings of a particular renderer or traversal: wavefront path tracing~\cite{laine2013wavefront}, persistent threads~\cite{aila2009gpurt, gupta2012persistent}, ray reordering~\cite{meister2020reordering, pharr1997memcoherent}, and, for memory coherence in traversals, packet tracing~\cite{wald2001packet} and treelets~\cite{aila2010treelet}.
Production renderers and traversal libraries~\cite{pbrt4, lee2017moonray, christensen2024rendermanxpu, wald2014embree, sawhney2021fcpw} compose several of these with tiling and vectorization, but each composition is written into the application and tied to one machine.

\textit{Compilers for divergent graphics workloads}.
The GPU coroutines of \textsc{Luisa}~\cite{zheng2024gpucoroutine} are the closest prior work to \schedule{defer}: suspension points annotated in a kernel are compiled into continuation frames, which one of a fixed set of schedulers (wavefront, persistent threads) executes.
Dr.Jit~\cite{jakob2022drjit} traces an SPMD program and executes the trace either fused into a megakernel or split at loops and virtual calls.
\textsc{Luisa}'s suspension points are baked into the kernel, so exploring a schedule means editing the program, and Dr.Jit's split is a mode of the JIT; neither is a decision about a particular call, and neither compiler exposes tiling, vectorization, or specialization to compose with deferral.
Rodent~\cite{perardgayot2019rodent} and RaTrace~\cite{perardgayot2017ratrace} obtain portable renderers and traversals by partially evaluating~\cite{leissa2018anydsl} a library of higher-order traversal and shading variants against a scene and target, so their execution strategies devolve to a choice of library functions, not a composable scheduling space; we do think their partial evaluation would be useful in \sys for further data-specific specialization.

\textit{Stage-graph runtimes, stream languages, and parallel IRs}.
GRAMPS~\cite{sugerman2009gramps}, Piko~\cite{patney2015piko}, and the GPU task runtimes of \citet{tzeng2010taskmanage}, Softshell~\cite{steinberger2012softshell}, and Whippletree~\cite{steinberger2014whippletree} schedule graphs of stages connected by queues with dynamic rates; GRAMPS inspired our treatment of data flow, and Piko goes furthest in separating a stage's logic from its scheduling (binning and ordering).
A StreamIt~\cite{thies2002streamit} program is the same kind of graph, which its compiler~\cite{gordon2002streamit} merges and splits to fit the machine, but a highly static one: stages push a fixed number of items per run, so queue expansion rates and loop counts are known at compile time; Brook~\cite{buck2004brook}, the predecessor to CUDA~\cite{lindholm2008tesla}, maps kernels over fixed-length streams and has no queues at all.
HPVM~\cite{kotsifakou2018hpvm}, a hierarchical dataflow graph with data-parallel nodes that targets CPUs, vector units, and GPUs, is the closest IR to a PDG, but is static in the same way: each edge carries one value per node instance, so a node cannot push to a queue on some instances and not others, and a drain cannot feed a queue upstream of itself.
In each of these systems, the programmer writes the stages and the enqueues; a PDG derives both from the control flow of an SPMD program, so moving a stage boundary is a change to the schedule.
Other parallel IRs target fork-join parallelism~\cite{schardl2019tapir, rainey2021parasm, srinivasan1993ssapar}; to the best of our knowledge, PDGs are the first parallel IR that models dynamic data flow alongside control flow and parallelism in one representation, and supports explicit transformation via a scheduling language.

\textit{User-schedulable languages}.
\sys's loop transformations are the standard loop transformations~\cite{bacon1994transforms} of array DSLs~\cite{chen2008chill, jrk2012halide, jrk2013halide, ikarashi2022exo, ikarashi2025exo2}, which schedule dense array programs whose loop structure is known at compile time, and whose schedules are pull-based: each producer is computed at some level of its consumer's loop nest.
Divergent programs instead discover work at runtime, so \sys's schedules are \textit{push}-based, driven by the producers whose work is deferred, and its new transformations rewrite control flow and data flow in addition to loops.
Sparse tensor and graph compilers~\cite{kjolstad2017taco, senanayake2020scheduling, strout2018spf, zhao2022spf, zhang2018graphit} schedule \textit{irregular} programs, and some of their directives similarly target load imbalance and memory divergence, but the irregularity they schedule is in the \textit{data}: index structures determine the iteration space, and a schedule transforms that iteration space.
The divergence \sys is designed for is also in the \textit{control}: every lane decides at runtime what execution path to take (e.g., how deeply to recurse) as a product of computation on the data, so its schedules must be dynamic and reactive.

\textit{SPMD compilation and vectorization}.
%
ispc~\cite{pharr2012ispc}, whole-function vectorization~\cite{karrenberg2011wholefuncvec}, and partial linearization~\cite{moll2018pcfl} map SPMD code onto SIMD units as SIMT hardware does~\cite{lindholm2008tesla}: one instance per lane, with control flow linearized under masks.
Our vectorization transformation builds directly on this work, including its notions of uniform and varying data and control flow.
However, these techniques are standalone, leaving programmers to compose vectorization with other scheduling transformations by hand, as the MoonRay~\cite{lee2017moonray} developers did.

\textit{Machine models and portability}.
Sequoia~\cite{fatahalian2006sequoia} separates an algorithm from its mapping onto an explicit hierarchy of memories, Exo~\cite{ikarashi2022exo} moves hardware definitions into user libraries, and Mosaic~\cite{bansal2023mosaic} supports accelerator binding intermixed with loop scheduling for sparse tensor algebra; our hardware models (\Cref{sec:hardware-models}) follow each of these, including support for graphics accelerators.

\textit{Continuations}.
\schedule{defer} is a selective CPS conversion~\cite{appel2007continuations} followed by a defunctionalization~\cite{reynolds1972defunc} specialized per producer~\cite{brandon2023defunc}.
The transformation is not particularly novel, but we believe the user control and composition with loop scheduling and hardware bindings is.

\section{Conclusion and Future Work}
\label{sec:conclusion}

Parallel dispatch graphs represent control flow, data flow, and parallel loops in the same representation, and enable composable scheduling transformations to map SPMD programs onto SIMD hardware.
The execution strategies of hand-optimized tree traversals, path tracers, and Monte Carlo PDE solvers are points in that scheduling space, and \sys matches or outperforms the hand-written originals.
Just as \citet{gyurgyik2026decoupling} show that the optimal data layout is (machine, algorithm, data)-dependent, we show the same is true for the execution \textit{schedule}.
%
However, while our evaluation holds tree data layouts constant, we note that layouts and schedules are necessarily interdependent:
vectorized tree traversals or queue drains work best when the corresponding data is stored in SoA, and some renderers argue for more complex queue layouts~\cite{lee2017moonray}.
%
Together, these observations expose an interesting direction for future work: profile-guided co-optimization of layouts and schedules, across both the trees being traversed and the intermediate work queues that wavefront schedules drain.
More generally, we believe that portable, high-performance code requires decoupling the program specification from the decisions that only affect performance (data structures, layouts, and schedules).
Decoupling alone does not provide performance, however; it must be paired with a compiler that specializes the program to those decisions.



\begin{acks}
%
%
We thank Rohan Yadav for answering so, so, so many questions about GPUs and the visualization of parallel kernels.
We also thank
Atharva Chougule,
James Dong,
Olivia Hsu,
Sang Yoon Kim,
Rubens Lacouture,
Katherine Mohr,
Sai Gautham Ravipati,
Dalia Shaaban,
Nathan Sobotka,
Rupanshu Soi,
Shiv Sundram,
Usman Tariq,
and Bobby Yan
for their helpful feedback on an early draft of this work.
Additionally, we thank
Sai Praveen Bangaru,
Fr\'edo Durand,
Aaron Lefohn,
Zander Majercik,
Brennan Shacklett,
and Venki Sivaram,
for discussion on this work.
Alexander and Christophe were supported by the Qualcomm Innovation Fellowship during part of this work.
Alexander was also supported by the NVIDIA Graduate Fellowship and took part in this work while interning at NVIDIA under Matt and Rohan.

\end{acks}

\bibliographystyle{ACM-Reference-Format}
\bibliography{references}

\newpage
\appendix 

\section{RTIOW Renderer Intermediate Representation}
\label{appendix:rtiow-ir}
\input{figures/lst-rtiow-ir}

\section{Defer Implementation}
\label{appendix:defer-impl}

\Cref{alg:defer} is the deferral transformation of \Cref{sec:transforms}; $\mathit{Defer}_q$ is the \code{Defer} arm for queue $q$, and \textsc{PlaceAtOwner} performs the legality check described in \Cref{sec:transforms}.

\input{figures/alg-defer}

\section{Dataset Statistics}
\label{sec:appendix-statistics}

\begin{table}[H]
\footnotesize
\caption{\footnotesize Meshes of the tree traversal (\Cref{sec:eval-tree-traversals}) and Monte Carlo PDE (\Cref{sec:eval-wosx}) benchmarks.
Meshes come from pbrt-v4-scenes~\cite{pbrt4} and the six demo meshes of the WoSX repository~\cite{sawhney2026wosx} ($\dagger$), used only for the PDE benchmarks. \Cref{fig:rtq} omits lte-orb and \Cref{fig:wosx} omits landscape, which are degenerate inputs for those benchmarks.}
\label{tab:meshes}
\begin{tabular}{@{}llr@{}}
\toprule
mesh & file & triangles \\
\midrule
comb$^\dagger$ & \texttt{electrostatics/data/mems-comb-electrodes.obj} & 600 \\
pavilion & \texttt{barcelona-pavilion/geometry/mesh\_00099.ply} & 11,366 \\
head & \texttt{head/geometry/head.ply} & 17,674 \\
zero-day & \texttt{zero-day/geometry/9dbbf607\ldots{}.ply} & 48,960 \\
whale$^\dagger$ & \texttt{potential\_flow/data/whale.obj} & 99,648 \\
bmw & \texttt{bmw-m6/geometry/mesh\_00037.ply} & 110,592 \\
crown & \texttt{crown/geometry/mesh\_00397.ply} & 155,520 \\
ivy & \texttt{sanmiguel/geometry/enredadera\_00001.ply} & 179,603 \\
car$^\dagger$ & \texttt{potential\_flow/data/car.obj} & 324,640 \\
villa & \texttt{villa/geometry/mesh\_00072.ply} & 390,784 \\
octopus$^\dagger$ & \texttt{geometric\_deformation/data/octopus.obj} & 497,236 \\
beast$^\dagger$ & \texttt{geometric\_deformation/data/beast.obj} & 606,136 \\
dambreak & \texttt{dambreak/geometry/dambreak1.ply} & 1,015,024 \\
sportscar & \texttt{sportscar/geometry/OSD\_CarPoly\_0000\_m028.ply} & 1,091,232 \\
rover$^\dagger$ & \texttt{thermal\_conduction/data/rover.zip} & 1,616,316 \\
landscape & \texttt{landscape/geometry/mesh\_00370.ply} & 1,916,928 \\
lte-orb & \texttt{lte-orb/geometry/mesh-3.ply.gz} & 3,423,232 \\
ganesha & \texttt{ganesha/geometry/ganesha.ply.gz} & 4,323,658 \\
dragon & \texttt{sssdragon/geometry/dragon.ply.gz} & 7,219,045 \\
\bottomrule
\end{tabular}
\end{table}

\begin{table}[H]
\footnotesize
\caption{\footnotesize PBRT evaluation scenes. Scenes named by filename in pbrt-v4-scenes~\cite{pbrt4}, except (for brevity) sanmiguel (sanmiguel/sanmiguel-courtyard), sanmiguel-2 (sanmiguel/sanmiguel-courtyard-second), bistro (bistro/bistro\_vespa), landscape (landscape/view-0), watercolor (watercolor/camera-1), kroken (kroken/camera-1).}
\label{tab:dataset-statistics}
\begin{tabular}{@{}lrrrrrrlr@{}}
\toprule
scene & triangles & meshes & instances & materials & textures & lights & resolution & depth \\
\midrule
clouds & 0 & 0 & -- & 1 & 0 & 1 & 1920$\times$1080 & 50 \\
bunny-cloud & 0 & 0 & -- & 2 & 0 & 1 & 1920$\times$1080 & 50 \\
disney-cloud & 0 & 0 & -- & 2 & 0 & 2 & 1280$\times$720 & 100 \\
explosion & 0 & 0 & -- & 2 & 0 & 1 & 1300$\times$1800 & 5 \\
plume & 2 & 1 & -- & 3 & 0 & 1 & 1200$\times$1200 & 5 \\
head & 17,674 & 1 & -- & 2 & 3 & 1 & 1920$\times$1080 & 2 \\
book & 21,246 & 3 & -- & 4 & 9 & 2 & 1920$\times$1080 & 5 \\
killeroo-simple & 66,532 & 4 & -- & 5 & 0 & 1 & 700$\times$700 & 5 \\
killeroo-coated-gold & 532,230 & 4 & -- & 5 & 6 & 2 & 1368$\times$1026 & 5 \\
killeroo-gold & 532,230 & 4 & -- & 5 & 6 & 2 & 1368$\times$1026 & 5 \\
lte-orb-rough-glass & 597,812 & 8 & -- & 5 & 6 & 2 & 1200$\times$1200 & 30 \\
lte-orb-silver & 597,812 & 8 & -- & 5 & 6 & 2 & 1200$\times$1200 & 5 \\
dambreak0 & 620,956 & 8 & -- & 3 & 0 & 1 & 1920$\times$1080 & 64 \\
bmw-m6 & 792,746 & 114 & -- & 29 & 0 & 1 & 1400$\times$1000 & 8 \\
frame542 & 1,317,820 & 229 & 5 & 2 & 0 & 1 & 1920$\times$840 & 256 \\
sanmiguel & 2,503,021 & 796 & 69,040 & 300 & 1,069 & 1 & 1600$\times$1100 & 5 \\
sanmiguel-2 & 2,503,044 & 808 & 159,924 & 300 & 1,069 & 1 & 1600$\times$1100 & 5 \\
villa-daylight & 2,624,966 & 358 & 146 & 69 & 159 & 1 & 1800$\times$870 & 10 \\
villa-lights-on & 2,624,966 & 358 & 146 & 69 & 159 & 8 & 1000$\times$444 & 5 \\
bistro\_cafe & 2,852,758 & 1,693 & -- & 133 & 409 & 20,639 & 1920$\times$1080 & 5 \\
bistro & 2,852,758 & 1,693 & -- & 133 & 409 & 20,639 & 1920$\times$1080 & 5 \\
crown & 3,540,226 & 798 & -- & 52 & 23 & 12 & 1000$\times$1400 & 100 \\
ganesha & 4,323,668 & 6 & -- & 2 & 3 & 5 & 1440$\times$1440 & 5 \\
lte-orb-simple-ball & 5,237,888 & 4 & -- & 3 & 0 & 1 & 1200$\times$1200 & 5 \\
pavilion-day & 5,311,719 & 120 & 43 & 39 & 127 & 1 & 1600$\times$850 & 5 \\
sportscar-sky & 7,157,264 & 48 & -- & 49 & 0 & 1 & 1920$\times$1080 & 5 \\
dragon\_10 & 7,219,557 & 2 & -- & 3 & 0 & 1 & 1366$\times$1024 & 5 \\
landscape & 27,885,845 & 370 & 30,034 & 638 & 1,611 & 1 & 3840$\times$2160 & 5 \\
watercolor & 58,383,675 & 944 & 54 & 261 & 412 & 2 & 1920$\times$1440 & 15 \\
kroken & 287,904,585 & 248 & 10 & 80 & 164 & 2,176 & 1500$\times$1500 & 15 \\
\bottomrule
\end{tabular}
\end{table}

\section{PBRT Compile Times}
\label{appendix:compile-times}

\begin{figure}[H]
    \centering
    \includegraphics[width=\linewidth]{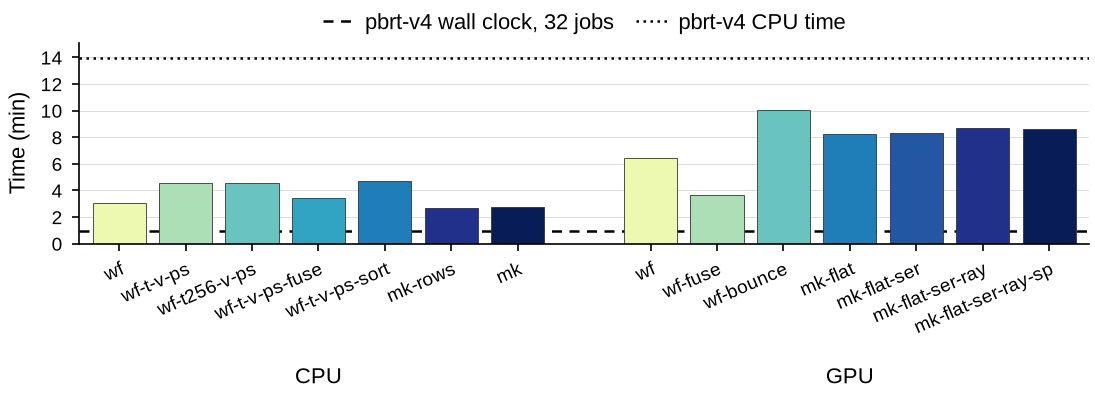}
    \caption{\sys compile times for the schedules used in \Cref{sec:eval-pbrt}. \sys compilation is not parallelized, but PBRT is also compiling features (e.g., an interactive viewport) that our implementation does not include.}
    \label{fig:pbrt-compile-times}
\end{figure}

\section{PBRT's Parallel Dispatch Graph}
\label{sec:appendix-pbrt-pdg}

\Cref{fig:pcg-pbrt-wavefront} shows PBRT's GPU wavefront integrator~\cite[Ch.~15.2]{pbrt4} as the PDG that \sys's 64-directive schedule produces; the graph is the body of one tile (\code{band}, \code{i}) of the pixel loop.
\texttt{camera\_ray} fills the ray queue \code{r}; \texttt{ray\_samples} draws each ray's samples, and the drain of \code{r} runs on OptiX threads~\cite{parker2010optix} and traces on the RT cores~\cite{nvidia2018turing}.
Every edge out of the dispatches that follow is deferred: misses go to \code{e}; rays inside a medium to \code{s}, whose kernel scatters into \code{S} or passes to the surface hit; and hits dispatch on the material, emissive hits into \code{a}, interface materials back into \code{r}, and the remaining 22 materials into the queues $m_1 \ldots m_n$ of one \schedule{specialize}, so each \texttt{bsdf} kernel is compiled for one material.
Subsurface materials defer again into \code{b}, whose probe rays \code{P} are traced on the RT cores and resumed by \texttt{sss}.
Every kernel that extends a path appends to \code{r} while \code{d < D}, and every kernel that samples a light appends to the shadow queue \code{w}, also traced on the RT cores.
\code{r} is the only queue its own drain can feed (the interface edge), so it is double-buffered, and the graph re-executes until \code{r} is empty (\Cref{sec:pdgs}).

\input{figures/fig-pcg-pbrt}

\section{Queue Size Comparison}
\label{appendix:queue-size}

\Cref{tab:queue-bytes} compares the bytes per stored state in PBRT's wavefront integrator and in \sys's matching schedule.
\sys's states are smaller for every queue but one, for two mechanical reasons: PBRT stores each boolean in 4 bytes, where \sys packs flags into the bits of one byte, and PBRT stores several 8-byte pointers (e.g., to a material) where \sys stores the 32-bit index of the same object in the per-material array.
The exception is BSSRDF evaluation, where PBRT's GPU integrator stores an approximation of the BSSRDF state that does not match the semantics of its CPU integrator; \sys, which reproduces the CPU semantics, carries 12 more bytes.

\begin{table}[H]
    \centering
    \begin{tabular}{lrr}
      \toprule
      Queue & PBRT & \sys \\
      \midrule
      Rays                        & 184 & 173 \\
      Medium samples              & 364 & 215 \\
      Medium scatters             & 120 & 112 \\
      Escaped rays                & 164 & 145 \\
      Emissive hits               & 192 & 169 \\
      Material hits (each of 22)  & 252 & 247 \\
      Shadow rays                 & 124 &  96 \\
      BSSRDF evaluation           & 168 & 180 \\
      Subsurface probes           & 308 & 184 \\
      Subsurface exits            & 308 & 220 \\
      Exit shadow rays            & 124 &  96 \\
      \bottomrule
    \end{tabular}
    \caption{Bytes per stored state in PBRT's wavefront integrator and in \sys's matching schedule. PBRT reuses one queue for the subsurface probes and exits, and its shadow queue for both shadow traces.}
    \label{tab:queue-bytes}
  \end{table}

\section{WoSX Schedules}
\label{sec:appendix-wosx-schedule}

\Cref{lst:wosx-schedule} is the schedule matching WoSX's persistent kernel.
The solver has two parallel loops, over points (\texttt{i}) and over walks (\texttt{w}); the schedule reorders them so walks are outermost, collapses them into a single loop over (walk, point) work items so that consecutive items are consecutive points (WoSX's order), and makes that loop persistent.
The tree traversals use the schedule of \Cref{sec:eval-tree-traversals}.

The other schedules of \Cref{fig:wosx} each change one or two scheduling commands.
\texttt{parallel} drops \texttt{persistent} and launches one thread per work item.
\texttt{persistent-reorder} collapses with points outermost instead, so that the lanes of a warp claim the walks of one point.
\texttt{persistent-256k} uses twice as many persistent threads, and \texttt{persistent-reorder-256k} does both.
The schedule \texttt{parallel-reorder-blk64} is a plain launch with points outermost and 64-thread blocks, so that inputs with few points still fill the GPU.
\texttt{optix-ser} binds the walks to OptiX threads and sorts the walk's recursive call by the Morton code of its position, which lowers to shader execution reordering~\cite{nvidia2022ser} before every step.
\texttt{defer-block} binds points to GPU blocks and defers each step of a walk into a per-point queue in shared memory, which is drained in rounds.

\vspace{0.75em}

\begin{lstlisting}[language=schedule, style=code, mathescape=true, basicstyle=\scriptsize\ttfamily, backgroundcolor=\color{codebackground}, frame=single, caption={Schedule matching WoSX's persistent GPU kernel (\texttt{persistent} in \Cref{fig:wosx}).},
label={lst:wosx-schedule}, aboveskip=0pt, belowskip=0pt]
solve_all.reorder(i, w)                  // walks outside points
k = solve_all.collapse(w, i)             // one work item per (walk, point) 
t = solve_all.persistent(k, 131072)      // 131072 workers claim items from a counter
solve_all.gpu_tile(t, 256)               // 256 threads per block
\end{lstlisting}

\end{document}

%% file: macros.tex
\newcommand{\sys}{\textsc{Grove}\xspace}
\newcommand{\bonsai}{\textsc{Bonsai}\xspace}

\definecolor{codebackground}{RGB}{252,252,252}

\definecolor{color0}{RGB}{255,255,217}
\definecolor{color1}{RGB}{237,248,177}
\definecolor{color2}{RGB}{199,233,180}
\definecolor{color3}{RGB}{127,205,187}
\definecolor{color4}{RGB}{65,182,196}
\definecolor{color5}{RGB}{29,145,192}
\definecolor{color6}{RGB}{34,94,168}
\definecolor{color7}{RGB}{37,52,148}
\definecolor{color8}{RGB}{8,29,88}

\definecolor{FireBrickRed}{RGB}{178,34,34}
\definecolor{DarkCyan}{RGB}{30,144,255}
\definecolor{LessDarkCyan}{RGB}{20,110,255}

\lstdefinelanguage{input}{
  keywords={func, parfor, if, return, for, in, else, sync, alloc, while, swap, match, switch, case, type, spawn, join, atomic, do},
  keywordstyle=\color{FireBrickRed}\textbf,
  identifierstyle=\color{black},
  sensitive=true,
  comment=[l]{//},
  morecomment=[s]{/*}{*/},
  commentstyle=\color{gray}\ttfamily,
  stringstyle=\color{red}\ttfamily,
  morekeywords = [2]{interact, Li, trace, background, render, dot, rngu, eval, ref, reflect, camera_ray, lambertian, metal, dielectric, light},
  keywordstyle = [2]\color{teal},
  morekeywords = [3]{append, empty},
  keywordstyle = [3]\color{purple},
  morekeywords = [4]{Lambertian, Metal, Dielectric, DiffuseLight, Absorb, Scatter, None, Some, Ray, Material, Response, Result, Done, Defer, Value, Index, mut, reduce},
  keywordstyle = [4]\color{LessDarkCyan},
}

\lstdefinelanguage{schedule}{
  keywords={schedule},
  keywordstyle=\color{FireBrickRed}\textbf,
  identifierstyle=\color{black},
  sensitive=true,
  comment=[l]{//},
  morecomment=[s]{/*}{*/},
  commentstyle=\color{gray}\ttfamily,
  stringstyle=\color{red}\ttfamily,
  morekeywords = [2]{queue, specialize, split, collapse, reorder, persistent, vectorize, loopify, defer, stage, specialize, sort, bind, prefetch, gpu_tile, unroll},
  keywordstyle = [2]\color{FireBrickRed},
  morekeywords = [3]{interact, Li, trace, background, render, dot, rngu, eval, ref, reflect, camera_ray, lambertian, metal, dielectric, light, triangles, root, cont},
  keywordstyle = [3]\color{teal},
  morekeywords = [4]{CPUThread, GPUBlock, GPUThread, OptixThread, RTCore, TextureUnit, Heap, Stack, Managed, Device, DeviceShared, Extern, ExternDevice, ExternManaged, ThreadLocal},
  keywordstyle = [4]\color{LessDarkCyan},
  morekeywords = [5]{Interior, Leaf, AABB, Ray},
  keywordstyle = [5]\color{brown},
}

\lstdefinelanguage{example}{
  keywords={func, if, type, match, with, else, while, for},
  keywordstyle=\color{red},
  identifierstyle=\color{black},
  sensitive=true,
  comment=[l]{//},
  morecomment=[s]{/*}{*/},
  commentstyle=\color{gray}\ttfamily,
  stringstyle=\color{red}\ttfamily,
  morekeywords = [2]{trace, intersect_triangle, intersect},
  keywordstyle = [2]\color{blue},
  morekeywords = [3]{BVH, Ray, Result, Triangle, AABB},
  keywordstyle = [3]\color{purple},
  morekeywords = [4]{Leaf, Node},
  keywordstyle = [4]\color{teal},
}

\lstdefinelanguage{terminators}{
  keywords={jmp, cjmp, call, ret, par, detach, yield, mcall, defer},
  keywordstyle=\color{FireBrickRed}\textbf,
  identifierstyle=\color{black},
  sensitive=false,
  comment=[l]{//},
  morecomment=[s]{/*}{*/},
  commentstyle=\color{gray}\ttfamily,
  stringstyle=\color{red}\ttfamily,
  morekeywords = [2]{f},
  keywordstyle = [2]\color{teal},
  morekeywords = [3]{block, block0, cont, body, block1},
  keywordstyle = [3]\color{brown},
}

\lstdefinelanguage{ir}{
  keywords={func, jmp, cjmp, call, ret, par, detach, yield, mcall, defer},
  keywordstyle=\color{FireBrickRed}\textbf,
  identifierstyle=\color{black},
  comment=[l]{//},
  morecomment=[s]{/*}{*/},
  commentstyle=\color{gray}\ttfamily,
  stringstyle=\color{red}\ttfamily,
  morekeywords = [2]{interact, Li, trace, background, render, dot, rngu, eval, ref, reflect, camera_ray, lambertian, dielectric},
  keywordstyle = [2]\color{teal},
  morekeywords = [3]{lamb, lamb_c, metal, metal_c, metal_cc, metal_ccc, metal_s, metal_a, diel, diel_c, light, light_c, scatter, li_tr, li_hit, li_miss, li_some, li_resp, li_emit, li_scatter, body, cont, body_store
  },
  keywordstyle = [3]\color{brown},
  morekeywords = [4]{Lambertian, Metal, Dielectric, DiffuseLight, Absorb, Scatter, None, Some, Ray, Material, Response, Result, Done, Defer, Value, Index, mut, reduce},
  keywordstyle = [4]\color{LessDarkCyan},
}

\lstdefinelanguage{psuedo}{
  keywords={func, while, if, do, else, for, return, swap},
  keywordstyle=\color{red},
  identifierstyle=\color{black},
  sensitive=false,
  comment=[l]{//},
  morecomment=[s]{/*}{*/},
  commentstyle=\color{gray}\ttfamily,
  stringstyle=\color{red}\ttfamily,
}

\lstdefinelanguage{cuda}{
  keywords={while, if, do, else, for, return, syncblock, threadIdx, shared, get_tid},
  keywordstyle=\color{red},
  identifierstyle=\color{black},
  comment=[l]{//},
  morecomment=[s]{/*}{*/},
  commentstyle=\color{gray}\ttfamily,
  stringstyle=\color{red}\ttfamily,
  morekeywords = [2]{void, Tree, Ray, float3, tuple, int, Intersection, auto, optional},
  keywordstyle = [2]\color{cyan},
  morekeywords = [3]{kernel, color, brdf_eval, sky_color},
  keywordstyle = [3]\color{blue},
  morekeywords = [4]{mq_miss_cont, hq_hit_cont},
  keywordstyle = [4]\color{purple},
}

\lstdefinelanguage{rwir}{
  keywords={while, if, do, else, for, return},
  keywordstyle=\color{red},
  identifierstyle=\color{black},
  comment=[l]{//},
  morecomment=[s]{/*}{*/},
  commentstyle=\color{gray}\ttfamily,
  stringstyle=\color{red}\ttfamily,
  morekeywords = [2]{void, list, auto, ordered_set, tuple},
  keywordstyle = [2]\color{cyan},
  morekeywords = [3]{IR, Cursor, Queue_t, Block, Index, Instruction, Terminator, Var, Type},
  keywordstyle = [3]\color{blue},
  morekeywords = [4]{defer, GetBlock, RewritePaths, MakeAllocate, MakeLoad, MakeCJmp, MakeYield, CountField, GetProducerTerm, MakePar, MakeTaggedUnion},
  keywordstyle = [4]\color{purple},
}

\def\irlang|#1|{\texttt{\lstinline[language=ir]|#1|}}
\def\inlang|#1|{\lstinline[language=input]|#1|}

\def\schedlang|#1|{\texttt{\lstinline[language=schedule]|#1|}}

\definecolor{myhighlightcolor}{rgb}{1.0, 0.9, 0.7} 

\newcommand{\hlrange}[2]{%
  \ifnum\value{lstnumber}<#1\else%
    \ifnum\value{lstnumber}>#2\else%
       \color{myhighlightcolor}%
    \fi%
  \fi%
}

\algnewcommand{\LComment}[1]{\State \textcolor{black!60}{\(\triangleright\) #1}}

\tikzset{
    diagram config/.style={
        node distance=0.5cm,
        font=\sffamily
    },
    sysbox/.style={
        rectangle, draw, rounded corners=1ex,
        minimum width=1.5cm, minimum height=0.5cm,
        fill=color4!50, text centered, thick,
        text height=1.5ex, text depth=0.5ex
    },
    sysqueue/.style={
        cylinder, draw, thick,
        shape border rotate=90, 
        minimum width=1cm, minimum height=0.5cm,
        fill=color6!50, text centered,
        aspect=0.3,
        text height=2ex, text depth=0.8ex, text width=3ex, align=center
    },
    arrow bold/.style={->, ultra thick, >=Latex},
    arrow regular/.style={->, thick, >=Latex},
    arrow dashed/.style={->, thick, dash pattern=on 3pt off 2pt, >=Latex},
    arrow dotted/.style={->, thick, dotted, >=Latex},
    bottom bold/.style={
        path picture={
            \draw[line width=1.5pt] 
            ([yshift=1.25pt]path picture bounding box.south west) -- 
            ([yshift=1.25pt]path picture bounding box.south east);
        }
    },
}

\makeatletter
\newcommand{\setcolumn}[2]{
    \expandafter\def\csname col@#1@x\endcsname{#2}
    \expandafter\def\csname col@#1@prev\endcsname{}
}

\newcommand{\additem}[4]{
    \edef\prevnode{\csname col@#2@prev\endcsname}
    \ifx\prevnode\empty
        \node[diagram config, #1, xshift=\csname col@#2@x\endcsname] (#3) {#4};
    \else
        \node[diagram config, #1, below=of \prevnode] (#3) {#4};
    \fi
    \expandafter\def\csname col@#2@prev\endcsname{#3}
}
\makeatother

\newlength{\downshift}
\tikzset{
  tablecell/.style={
    draw,
    minimum width=2cm,
    minimum height=0.5cm,
    anchor=center,
    text height=1.5ex,
    text depth=0.25ex
  },
  masked/.style={
    fill=gray!20!white,
    text=gray!50,
    postaction={draw, gray, thin, shorten >=2pt, shorten <=2pt, -} 
  }
}

\def\emptycell{}

\def\colwidth{2cm}
\def\cellheight{0.5cm}

\newcommand{\drawcolumn}[3][1.5cm]{%
    \foreach \item [count=\i from 0] in {#3} {
        \foreach \name [expand list=true] / \customcolor in {\item/DEFAULT_MARKER} {
            \ifx\name\emptycell
                \def\cellcol{white}
                \def\showname{} 
            \else
                \def\showname{\name}
                \def\tempmarker{DEFAULT_MARKER}
                \ifx\customcolor\tempmarker
                    \def\cellcol{color4!50} 
                \else
                    \def\cellcol{\customcolor}
                \fi
            \fi
            
            \node[tablecell, 
                  fill=\cellcol, 
                  minimum width=#1, 
                  anchor=north west, 
                  outer sep=0pt] 
                at ({#2 * #1}, {-\i * \cellheight}) {\showname};
                
            \breakforeach
        }
    }
}

\newcommand{\drawrow}[3][1.5cm]{%
    \foreach \item [count=\i from 0] in {#3} {
        \foreach \name [expand list=true] / \customcolor in {\item/DEFAULT_MARKER} {
            \ifx\name\emptycell
                \def\cellcol{white}
                \def\showname{} 
            \else
                \def\showname{\name}
                \def\tempmarker{DEFAULT_MARKER}
                \ifx\customcolor\tempmarker
                    \def\cellcol{color4!50} 
                \else
                    \def\cellcol{\customcolor}
                \fi
            \fi
            
            \node[tablecell, 
                  fill=\cellcol, 
                  minimum width=#1, 
                  anchor=north west, 
                  outer sep=0pt] 
                at ({\i * #1}, {-#2 * \cellheight}) {\showname};
                
            \breakforeach
        }
    }
}

\newcommand{\addcontainer}[5][color2!50]{%
    \edef\prevnode{\csname col@#2@prev\endcsname}
    \ifx\prevnode\empty\else
        \node[coordinate, below=0.75cm of \prevnode] (#3_dummy) {};
        \expandafter\def\csname col@#2@prev\endcsname{#3_dummy}
    \fi
    \begin{scope}[local bounding box=#3_inner]
        #5
    \end{scope}%
    \begin{pgfonlayer}{background0}
        \node[sysbox, fill=#1,
              fit={([xshift=-1.2cm, yshift=1.5em]\csname col@#2@x\endcsname, 0 |- #3_inner.north)
                   ([xshift=1.2cm, yshift=-1em]\csname col@#2@x\endcsname, 0 |- #3_inner.south)},
              inner sep=0pt] (#3) {};
    \end{pgfonlayer}%
    \node[anchor=north west, font=\sffamily\bfseries, inner sep=6pt] at (#3.north west) {#4};%
    \expandafter\def\csname col@#2@prev\endcsname{#3}%
}

\newcommand{\addbginner}[4][]{
    \begin{pgfonlayer}{background1}
        \node[
            rectangle, 
            fill=#3, 
            rounded corners=2ex, 
            fit=#4, 
            inner xsep=0.1cm,
            inner ysep=0.45cm
        ] (#2) {};
        
        \def\bglbl{#1}
        \ifx\bglbl\empty\else
            \node[
                anchor=south, 
                inner sep=0.1cm,   
                font=\sffamily\small,
                align=center
            ] at (#2.south) {#1}; 
        \fi
    \end{pgfonlayer}
}

\newcommand{\addbgouter}[4][]{
    \begin{pgfonlayer}{background2}
        \node[
            rectangle, 
            fill=#3, 
            rounded corners=2ex, 
            fit=#4, 
            inner xsep=0.1cm,
            inner ysep=0.45cm
        ] (#2) {};
        
        \def\bglbl{#1}
        \ifx\bglbl\empty\else
            \node[
                anchor=south, 
                inner sep=0.1cm,   
                font=\sffamily\small,
                align=center
            ] at (#2.south) {#1}; 
        \fi
    \end{pgfonlayer}
}

\input{figures/pcg-styles}

\newcommand{\inputcode}[1]{\texttt{\lstinline[language=input]|#1|}}

\newlength{\cellw}
\newcount\nrows
\newif\ifcellempty
\def\splitcell#1/#2/#3\splitend{\def\name{#1}\def\customcolor{#2}}
\newcommand{\parsecell}[1]{%
  \expandafter\splitcell#1//\splitend
  \cellemptyfalse
  \ifx\name\empty\cellemptytrue\else\ifx\name\emptycell\cellemptytrue\fi\fi}

\newcommand{\tracerow}[1]{%
  \expandafter\gdef\csname rowdata\the\nrows\endcsname{#1}%
  \global\advance\nrows 1 }

\newcommand{\measurerow}[1]{%
  \foreach \item [count=\i from 0] in {#1} {%
    \parsecell{\item}%
    \ifcellempty\else
      \pgfinterruptpicture
        \settowidth{\cellw}{\tikz\node[tablecell, draw=none, fill=none,
                                minimum width=0pt, outer sep=0pt] {\name};}%
        \ifdim\cellw>\csname colw\i\endcsname\relax
          \expandafter\xdef\csname colw\i\endcsname{\the\cellw}%
        \fi
      \endpgfinterruptpicture
    \fi}}

\newcommand{\drawtracerow}[1]{%
  \coordinate (cursor) at (0, {-\r*\cellheight});
  \foreach \item [count=\i from 0] in {#1} {%
    \parsecell{\item}%
    \ifcellempty
      \def\cellcol{white}\def\showname{}%
    \else
      \def\showname{\name}%
      \ifx\customcolor\empty\def\cellcol{color4!50}\else\def\cellcol{\customcolor}\fi
    \fi
    \edef\colwidth{\csname colw\i\endcsname}%
    \node[tablecell, fill=\cellcol, minimum width=\colwidth,
          anchor=north west, outer sep=0pt] (c-\r-\i) at (cursor) {\showname};
    \coordinate (cursor) at (c-\r-\i.north east);
  }}

\newcommand{\drawtrace}[1][0pt]{%
  \edef\lastrow{\the\numexpr\nrows-1\relax}%
  \foreach \c in {0,...,49} {\expandafter\xdef\csname colw\c\endcsname{#1}}%
  \foreach \r in {0,...,\lastrow} {%
    \expandafter\let\expandafter\thisrow\csname rowdata\r\endcsname
    \expandafter\measurerow\expandafter{\thisrow}}%
  \foreach \r in {0,...,\lastrow} {%
    \expandafter\let\expandafter\thisrow\csname rowdata\r\endcsname
    \expandafter\drawtracerow\expandafter{\thisrow}}%
  \global\nrows=0 }

\newcommand{\code}[1]{\texttt{\lstinline[language=input]|#1|}}

\newcommand{\algcomment}[1]{\textcolor{darkgray}{#1}}

\newcommand{\schedule}[1]{\texttt{\lstinline[language=schedule]|#1|}}

\colorlet{codehighlight}{yellow!30} 
\newcommand{\hlline}{\rlap{\color{codehighlight}%
  \kern-3pt\rule[-\dp\strutbox]{\dimexpr\linewidth+6pt\relax}{\baselineskip}}}

\lstdefinestyle{code}{
  mathescape=true,
  basicstyle=\scriptsize\ttfamily,
  backgroundcolor=\color{codebackground},
  frame=single,
  aboveskip=0pt,
  belowskip=0pt,
  abovecaptionskip=2pt,
  belowcaptionskip=0pt,
  captionpos=b,
  xleftmargin=3.4pt,
  xrightmargin=3.4pt,
}

\lstdefinestyle{syscode}{
  style=code,        
  language=input,
}

\lstdefinestyle{halfcode}{
  mathescape=true,
  basicstyle=\scriptsize\ttfamily,
  backgroundcolor=\color{codebackground},
  frame=single,
  aboveskip=0pt,
  belowskip=0pt,
  abovecaptionskip=2pt,
  belowcaptionskip=0pt,
  captionpos=b,
  xleftmargin=1.7pt,
  xrightmargin=1.7pt,
}

\lstdefinelanguage{bonsai}{
  keywords={from},
  keywordstyle=\color{FireBrickRed},
  identifierstyle=\color{black},
  sensitive=true,
  comment=[l]{//},
  morecomment=[s]{/*}{*/},
  commentstyle=\color{gray}\ttfamily,
  stringstyle=\color{red}\ttfamily,
}

\newcommand{\bonsaicode}[1]{\texttt{\lstinline[language=bonsai]|#1|}}

%% file: figures/pcg-styles.tex
\usepackage{graphicx}
\usetikzlibrary{arrows.meta, positioning, fit, backgrounds, shapes.geometric, shapes.multipart, calc}
\makeatletter
\newdimen\pcg@dima \newdimen\pcg@dimb \newdimen\pcg@leftneed \newdimen\pcg@maxlanex \newdimen\pcg@minlanex
\newcount\pcg@nrecur
\def\pcg@autoname{auto}\def\pcg@retname{ret}\def\pcg@selfname{self}\def\pcg@centername{center}
\def\pcg@v{v}\def\pcg@h{h}\def\pcg@hv{hv}
\def\pcg@deferred{}
\def\pcg@measure#1#2{\pgfinterruptpicture\settowidth{#1}{#2}\global#1=#1\endpgfinterruptpicture}

\tikzset{
  pcg/.style={font=\scriptsize\ttfamily, >={Stealth[length=4pt, width=3pt]}, line width=0.5pt,
              execute at begin picture={\pgfsetlayers{pcgbackdrop,background,main}}},
  backdrop/.style={fill=black!8, draw=none, rounded corners=3pt, inner sep=2mm},
  backdrop name/.style={font=\scriptsize\sffamily, text=black!70, anchor=south west, inner sep=2pt},
  enclosure/.style={draw=black!60, dotted, rounded corners=3pt, inner sep=2.5mm},
  enclosure name/.style={font=\scriptsize\sffamily, text=black!70, anchor=north west, inner sep=2pt},
  note/.style={font=\scriptsize\sffamily, text=black!70, fill=black!8, rounded corners=2pt, inner sep=2pt},
  box/.style={draw=color8, text=color8, rounded corners=2pt, align=center,
              inner xsep=4pt, inner ysep=2.5pt, minimum height=13pt},
  op/.style={box, fill=color4!65},
  cond/.style={box, fill=color2, font=\tiny\ttfamily, minimum height=10pt},
  dispatch/.style={box, rectangle split, rectangle split parts=2,
                   rectangle split part fill={color4!65, color2},
                   every two node part/.style={font=\tiny\ttfamily}},
  queue/.style={cylinder, shape border rotate=90, aspect=0.3, draw=color8, fill=color6,
                text=white, inner xsep=5pt, inner ysep=2pt, minimum height=16pt},
  fn/.style={draw=color8, fill=color0, rounded corners=3pt, inner sep=5pt,
             append after command={([xshift=3pt]\tikzlastnode.south west)
                                   edge[fn exit] ([xshift=-3pt]\tikzlastnode.south east)}},
  fn exit/.style={draw=color8, line width=1.3pt},
  fn name/.style={font=\scriptsize\bfseries\ttfamily, text=color8, anchor=north west, inner sep=2pt},
  call/.style={fn, text=color8, align=center, inner xsep=4pt, inner ysep=2.5pt, minimum height=13pt},
  flow/.style={->, draw=color8},
  branch/.style={->, draw=color8, dash pattern=on 2pt off 1.5pt},
  guard/.style={font=\tiny\ttfamily, text=color8, inner sep=1.5pt},
}
\pgfkeys{/pcg/.is family, /pcg,
  lane/.store in=\pcg@lane, route/.store in=\pcg@route, label/.store in=\pcg@labelopts,
  drop/.store in=\pcg@drop, entry/.store in=\pcg@entrycol,
  arm labels/.store in=\pcg@armlabels,    
  pad left/.store in=\pcg@padleft, pad right/.store in=\pcg@padright,
  pad bottom/.store in=\pcg@padbottom, pad top/.store in=\pcg@padtop,
  height/.store in=\pcg@height, height=,  
  match/.store in=\pcg@match, match=,     
  symmetric/.is if=pcg@sym,               
  lane margin/.store in=\pcg@lanemargin,     lane margin=3mm,     
  label margin/.store in=\pcg@labelmargin,   label margin=1.5mm,  
  name clearance/.store in=\pcg@nameclear,   name clearance=8mm,  
  loop gap/.store in=\pcg@loopgap,           loop gap=2.5mm,      
  loop step/.store in=\pcg@loopstep,         loop step=2mm,       
  arm step/.store in=\pcg@armstep,           arm step=2mm,        
  enter length/.store in=\pcg@enterlen,      enter length=7mm,    
  text size/.store in=\pcg@textsize,         text size=7pt,       
  pad bottom=4mm, pad top=2.5mm,
  tight/.style={},                                                
  loose/.style={lane margin=4mm, label margin=2mm, name clearance=9mm, loop gap=3mm,
                loop step=2.5mm, arm step=2.5mm, enter length=10mm, pad bottom=9mm, pad top=3mm}}
\newcommand{\pcgset}[1]{\pgfkeys{/pcg/.cd, #1}}

\pgfdeclarelayer{pcgbackdrop}
\newsavebox{\pcg@box}
\newif\ifpcg@fit \newif\ifpcg@quiet \newif\ifpcg@infn \newif\ifpcg@sym
\pgfkeys{/pcg/fit/.is if=pcg@fit, /pcg/quiet/.is if=pcg@quiet}
\newenvironment{pcg}[1][]{%
  \pgfmathsetmacro{\pcg@scale}{\pcg@textsize/7}
  \pgfmathsetlengthmacro{\pcg@lw}{0.5pt*\pcg@scale}%
  \pgfmathsetlengthmacro{\pcg@al}{4pt*\pcg@scale}\pgfmathsetlengthmacro{\pcg@aw}{3pt*\pcg@scale}%
  \begin{lrbox}{\pcg@box}\begin{tikzpicture}[pcg, scale=\pcg@scale, transform shape, line width=\pcg@lw,
      >={Stealth[length=\pcg@al, width=\pcg@aw]}, x=20mm, y=-10mm, #1]}%
 {\end{tikzpicture}\end{lrbox}%
  \pcg@dima=\ht\pcg@box \advance\pcg@dima by \dp\pcg@box
  \pgfmathsetmacro{\pcg@wmm}{int(round(\wd\pcg@box/2.845274))}%
  \pgfmathsetmacro{\pcg@hmm}{int(round(\pcg@dima/2.845274))}%
  \pgfmathsetmacro{\pcg@lmm}{int(round(\linewidth/2.845274))}%
  \typeout{pcg: picture \pcg@wmm mm x \pcg@hmm mm, line is \pcg@lmm mm wide}%
  \ifdim\wd\pcg@box>\linewidth
    \ifpcg@fit \resizebox{\linewidth}{!}{\usebox{\pcg@box}}%
    \else \PackageWarning{pcg}{picture is \pcg@wmm mm wide but the line is \pcg@lmm mm: it will
      overhang. Tighten it (x=, pads, lanes) or use \string\pcgset{fit}}%
      \ifpcg@quiet\else 
        \makebox[0pt][l]{\raisebox{\dp\pcg@box}[0pt][0pt]{\tiny\sffamily\color{red}%
          pcg: \pcg@wmm mm wide, line is \pcg@lmm mm}}\fi
      \usebox{\pcg@box}\fi
  \else \usebox{\pcg@box}\fi}

\newenvironment{fn}[3][]{%
  \pgfkeys{/pcg/.cd, entry=0, pad left=auto, pad right=auto, height=, match=, symmetric=false, #1}%
  \xdef\pcg@fn{#2}\gdef\pcg@fnname{#3}\xdef\pcg@fnentry{\pcg@entrycol}%
  \global\pcg@infntrue
  \global\pcg@nrecur=0 \global\pcg@leftneed=0pt
  \global\pcg@maxlanex=-\maxdimen \global\pcg@minlanex=\maxdimen
  \gdef\pcg@deferred{}%
  \begin{scope}[local bounding box=#2-body]%
}{%
  \end{scope}%
  \pcg@drawbox
  \pcg@deferred\gdef\pcg@deferred{}%
  \global\pcg@infnfalse
}

\def\pcg@sethgt#1#2{\expandafter\gdef\csname pcg@h@#1\endcsname{#2}}
\newif\ifpcg@rerun
\AtEndDocument{\ifpcg@rerun\PackageWarningNoLine{pcg}{box heights changed; rerun LaTeX to match them}\fi}
\def\pcg@drawbox{%
  \pgfpointanchor{\pcg@fn-body}{north}\pcg@dimb=\pgf@y
  \pgfpointanchor{\pcg@fn-body}{south}\advance\pcg@dimb by -\pgf@y 
  \pgfmathsetlength{\pcg@dima}{(\pcg@dimb+\pcg@padtop+\pcg@padbottom+10pt)*\pcg@scale}%
  \edef\pcg@natural{\the\pcg@dima}%
  \ifcsname pcg@h@\pcg@fn\endcsname
    \pcg@dima=\csname pcg@h@\pcg@fn\endcsname\relax \advance\pcg@dima by -\pcg@natural\relax
    \ifdim\pcg@dima<0pt \pcg@dima=-\pcg@dima\fi
    \ifdim\pcg@dima>0.5pt \global\pcg@reruntrue\fi
  \else \global\pcg@reruntrue\fi
  \protected@write\@auxout{}{\string\pcg@sethgt{\pcg@fn}{\pcg@natural}}%
  \ifx\pcg@match\@empty\else 
    \pcg@dima=\pcg@natural\relax
    \@for\pcg@m:=\pcg@match\do{%
      \ifcsname pcg@h@\pcg@m\endcsname
        \ifdim\csname pcg@h@\pcg@m\endcsname>\pcg@dima \pcg@dima=\csname pcg@h@\pcg@m\endcsname\relax\fi
      \else \global\pcg@reruntrue\fi}%
    \edef\pcg@height{\the\pcg@dima}%
  \fi
  \ifx\pcg@height\@empty\else 
    \pgfmathsetlength{\pcg@dima}{\pcg@height/\pcg@scale-\pcg@dimb-\pcg@padtop-10pt}
    \ifdim\pcg@dima<2mm
      \PackageWarning{pcg}{height=\pcg@height\space is too small for box \pcg@fn; using the smallest pad}%
      \pcg@dima=2mm\relax\fi
    \edef\pcg@padbottom{\the\pcg@dima}%
  \fi
  \ifx\pcg@padleft\pcg@autoname
    \pcg@dima=\pcg@leftneed
    \ifnum\pcg@nrecur>0 
      \pcg@measure{\pcg@dimb}{\scriptsize\bfseries\ttfamily\pcg@fnname}%
      \advance\pcg@dimb by \pcg@nameclear\relax
      \ifx\pcg@fnentry\pcg@centername \pgfpointanchor{\pcg@fn-body}{center}\else \pgfpointxy{\pcg@fnentry}{0}\fi
      \advance\pcg@dimb by -\pgf@x
      \pgfpointanchor{\pcg@fn-body}{west}\advance\pcg@dimb by \pgf@x
      \ifdim\pcg@dimb>\pcg@dima \pcg@dima=\pcg@dimb\fi
    \fi
    \ifdim\pcg@minlanex<\maxdimen 
      \pgfpointanchor{\pcg@fn-body}{west}\pcg@dimb=\pgf@x
      \advance\pcg@dimb by -\pcg@minlanex \advance\pcg@dimb by \pcg@lanemargin\relax
      \ifdim\pcg@dimb>\pcg@dima \pcg@dima=\pcg@dimb\fi
    \fi
    \ifdim\pcg@dima<0pt \pcg@dima=0pt\fi
  \else \pcg@dima=\pcg@padleft\relax \fi
  \edef\pcg@L{\the\pcg@dima}%
  \ifx\pcg@padright\pcg@autoname
    \pcg@dima=0pt
    \ifdim\pcg@maxlanex>-\maxdimen
      \pcg@dimb=\pcg@maxlanex \advance\pcg@dimb by \pcg@lanemargin\relax
      \pgfpointanchor{\pcg@fn-body}{east}\advance\pcg@dimb by -\pgf@x
      \ifdim\pcg@dimb>\pcg@dima \pcg@dima=\pcg@dimb\fi
    \fi
  \else \pcg@dima=\pcg@padright\relax \fi
  \edef\pcg@R{\the\pcg@dima}%
  \ifpcg@sym 
    \pcg@dima=\pcg@L\relax \pcg@dimb=\pcg@R\relax
    \ifdim\pcg@dima<\pcg@dimb \edef\pcg@L{\pcg@R}\else \edef\pcg@R{\pcg@L}\fi
  \fi
  \begin{scope}[on background layer]
    \node[fn, fit={(\pcg@fn-body) ([xshift=-\pcg@L]\pcg@fn-body.west) ([xshift=\pcg@R]\pcg@fn-body.east)
                   ([yshift=-\pcg@padbottom]\pcg@fn-body.south) ([yshift=\pcg@padtop]\pcg@fn-body.north)},
          label={[fn name]north west:\pcg@fnname}] (\pcg@fn) {};
  \end{scope}%
  \ifx\pcg@fnentry\pcg@centername \coordinate (\pcg@fn-entry) at (\pcg@fn.north);
  \else \coordinate (\pcg@fn-entry) at (\pcg@fn.north -| \pcg@fnentry,0);\fi
  \pgfpointanchor{\pcg@fn}{east}\pcg@dima=\pgf@x \pgfpointanchor{\pcg@fn}{west}\advance\pcg@dima by -\pgf@x
  \pgfpointanchor{\pcg@fn}{north}\pcg@dimb=\pgf@y \pgfpointanchor{\pcg@fn}{south}\advance\pcg@dimb by -\pgf@y
  \pgfmathparse{\pcg@scale*\pcg@dima/2.845274}\pgfmathprintnumberto[fixed, precision=1, assume math mode=true]{\pgfmathresult}{\pcg@bw}%
  \pgfmathparse{\pcg@scale*\pcg@dimb/2.845274}\pgfmathprintnumberto[fixed, precision=1, assume math mode=true]{\pgfmathresult}{\pcg@bh}%
  \typeout{pcg: box \pcg@fn\space is \pcg@bw mm x \pcg@bh mm}}

\def\pcg@node#1#2#3#4#5{\ifx\relax#4\relax\node[#1, #2] (#3) {#5};\else\node[#1, #2] (#3) at (#4) {#5};\fi}
\newcommand{\op}[4][]{\pcg@node{op}{#1}{#2}{#3}{#4}}
\newcommand{\cond}[4][]{\pcg@node{cond}{#1}{#2}{#3}{#4}}
\newcommand{\call}[4][]{\pcg@node{call}{#1}{#2}{#3}{#4}}
\newcommand{\dispatch}[5][]{\pcg@node{dispatch, yshift=5pt}{#1}{#2}{#3}{#4\nodepart{two}#5}%
  \expandafter\gdef\csname pcg@disp@#2\endcsname{}}
\def\pcg@startside#1#2{\ifcsname pcg@disp@#1\endcsname\edef\pcg@start{#1.two #2}\else\edef\pcg@start{#1.#2}\fi}

\newcommand{\flow}[3][]{\pcg@edge{flow}{#1}{#2}{}{#3}}
\newcommand{\case}[4][]{\pcg@edge{branch}{#1}{#2}{#3}{#4}}
\def\pcg@edge#1#2#3#4#5{%
  \pgfkeys{/pcg/.cd, lane=, route=auto, label=, drop=3mm, #2}%
  \def\pcg@tgt{#5}%
  \ifx\pcg@tgt\pcg@retname
    \ifx\pcg@lane\@empty
      \edef\pcg@add{\noexpand\pcg@doretdown{#1}{#3}{\unexpanded{#4}}{\unexpanded\expandafter{\pcg@labelopts}}}%
    \else
      \pgfpointxy{\pcg@lane}{0}%
      \ifdim\pgf@x>\pcg@maxlanex \global\pcg@maxlanex=\pgf@x\fi
      \ifdim\pgf@x<\pcg@minlanex \global\pcg@minlanex=\pgf@x\fi
      \edef\pcg@add{\noexpand\pcg@doretlane{#1}{#3}{\unexpanded{#4}}{\pcg@lane}{\unexpanded\expandafter{\pcg@labelopts}}}%
    \fi
  \else\ifx\pcg@tgt\pcg@selfname
    \global\advance\pcg@nrecur 1
    \ifx\pcg@lane\@empty \edef\pcg@lane{\the\pcg@nrecur}\fi
    \ifx\pcg@route\pcg@v 
      \edef\pcg@add{\noexpand\pcg@doselfdown{#1}{#3}{\unexpanded{#4}}{\pcg@lane}{\pcg@drop}{\unexpanded\expandafter{\pcg@labelopts}}}%
    \else 
      \pcg@measure{\pcg@dima}{\tiny\ttfamily #4}\advance\pcg@dima by \pcg@labelmargin\relax
      \ifdim\pcg@dima>\pcg@leftneed \global\pcg@leftneed=\pcg@dima\fi
      \edef\pcg@add{\noexpand\pcg@doself{#1}{#3}{\unexpanded{#4}}{\pcg@lane}{\unexpanded\expandafter{\pcg@labelopts}}}%
    \fi
  \else\ifx\pcg@lane\@empty
    \edef\pcg@add{\noexpand\pcg@doedge{#1}{#3}{\unexpanded{#4}}{#5}{\pcg@route}{\unexpanded\expandafter{\pcg@labelopts}}}%
  \else
    \edef\pcg@add{\noexpand\pcg@dolane{#1}{#3}{\unexpanded{#4}}{\pcg@lane}{#5}{\pcg@drop}{\unexpanded\expandafter{\pcg@labelopts}}}%
  \fi\fi\fi
  \ifpcg@infn \expandafter\g@addto@macro\expandafter\pcg@deferred\expandafter{\pcg@add}%
  \else \pcg@add \fi}

\def\pcg@doretdown#1#2#3#4{%
  \draw[#1] (#2) -- (#2 |- \pcg@fn.south) node[guard, midway, auto, #4] {#3};}
\def\pcg@laneside#1#2{%
  \pgfpointxy{#2}{0}\pcg@dima=\pgf@x \pgfpointanchor{#1}{center}%
  \ifdim\pcg@dima<\pgf@x \pcg@startside{#1}{west}\def\pcg@lanch{south east}\def\pcg@lsh{-1mm}%
  \else \pcg@startside{#1}{east}\def\pcg@lanch{south west}\def\pcg@lsh{1mm}\fi}
\def\pcg@doretlane#1#2#3#4#5{%
  \pcg@laneside{#2}{#4}%
  \draw[#1, rounded corners=2pt] (\pcg@start) -| (#4,0 |- \pcg@fn.south)
     node[guard, pos=0, anchor=\pcg@lanch, xshift=\pcg@lsh, #5] {#3};}
\def\pcg@dolane#1#2#3#4#5#6#7{%
  \pgfpointxy{#4}{0}\pcg@dima=\pgf@x \pgfpointanchor{#2}{center}%
  \ifdim\pcg@dima<\pgf@x \def\pcg@f{0.3}\def\pcg@lside{left}\else\def\pcg@f{0.7}\def\pcg@lside{right}\fi
  \pgfpointanchor{#5}{center}\ifdim\pcg@dima<\pgf@x \def\pcg@tside{west}\else\def\pcg@tside{east}\fi
  \coordinate (pcglane) at ($(#2.south west)!\pcg@f!(#2.south east)$);
  \draw[#1, rounded corners=2pt] (pcglane) -- ([yshift=-#6]pcglane |- \pcg@fn.south)
     node[guard, midway, \pcg@lside, #7] {#3}
     -| (#4,0 |- #5.center) -- (#5.\pcg@tside);}

\def\pcg@doself#1#2#3#4#5{%
  \pcg@startside{#2}{west}%
  \pgfmathsetlengthmacro{\pcg@ld}{\pcg@loopgap+\pcg@loopstep*(#4-1)}%
  \pgfmathsetlengthmacro{\pcg@lx}{-8mm+\pcg@loopstep*(#4-1)}%
  \draw[#1, rounded corners=2pt] (\pcg@start) -- (\pcg@start -| \pcg@fn.west)
     node[guard, pos=0, anchor=south east, xshift=-0.5mm, #5] {#3}
     -- ++(-\pcg@ld,0) |- ($(\pcg@fn-entry) + (\pcg@lx,\pcg@ld)$)
     -- ($(\pcg@fn-entry) + (\pcg@lx,0)$);}
\def\pcg@doselfdown#1#2#3#4#5#6{%
  \pgfmathsetlengthmacro{\pcg@ld}{\pcg@loopgap+\pcg@loopstep*(#4-1)}%
  \pgfmathsetlengthmacro{\pcg@lx}{-8mm+\pcg@loopstep*(#4-1)}%
  \coordinate (pcglane) at ($(#2.south west)!0.3!(#2.south east)$);
  \coordinate (pcgA) at ([xshift=-\pcg@ld]\pcg@fn.west);
  \coordinate (pcgB) at ([yshift=\pcg@ld]\pcg@fn.north);
  \coordinate (pcgL) at ($(\pcg@fn-entry) + (\pcg@lx,0)$);
  \draw[#1, rounded corners=2pt] (pcglane) -- ([yshift=-#5]pcglane |- \pcg@fn.south)
     node[guard, midway, left, #6] {#3}
     -| (pcgA |- pcgB) -| (pcgL);}
\def\pcg@doedge#1#2#3#4#5#6{%
  \pgfpointdiff{\pgfpointanchor{#2}{center}}{\pgfpointanchor{#4}{center}}%
  \pcg@dima=\pgf@x \pcg@dimb=\pgf@y
  \def\pcg@r{#5}%
  \ifx\pcg@r\pcg@autoname
    \ifdim\pcg@dima<1mm \ifdim\pcg@dima>-1mm \def\pcg@r{v}\fi\fi
    \ifx\pcg@r\pcg@autoname \ifdim\pcg@dimb<3mm \ifdim\pcg@dimb>-3mm \def\pcg@r{h}\fi\fi\fi
    \ifx\pcg@r\pcg@autoname \def\pcg@r{hv}\fi
  \fi
  \ifdim\pcg@dima<0pt
    \def\pcg@side{west}\def\pcg@oside{east}\def\pcg@lanch{south east}\def\pcg@lsh{-1mm}%
  \else
    \def\pcg@side{east}\def\pcg@oside{west}\def\pcg@lanch{south west}\def\pcg@lsh{1mm}%
  \fi
  \pcg@startside{#2}{\pcg@side}%
  \ifx\pcg@r\pcg@v
    \draw[#1] (#2) -- (#4) node[guard, midway, auto, #6] {#3};
  \else\ifx\pcg@r\pcg@h
    \draw[#1] (\pcg@start) -- (\pcg@start -| #4.\pcg@oside) node[guard, midway, above, #6] {#3};
  \else\ifx\pcg@r\pcg@hv
    \draw[#1, rounded corners=2pt] (\pcg@start) -| (#4)
       node[guard, pos=0, anchor=\pcg@lanch, xshift=\pcg@lsh, #6] {#3};
  \else
    \draw[#1, rounded corners=2pt] (#2) |- (#4) node[guard, pos=0.25, auto, #6] {#3};
  \fi\fi\fi}

\newcommand{\arms}[3][]{%
  \pgfkeys{/pcg/.cd, drop=4mm, arm labels=above, #1}%
  \foreach \pcgcase/\pcgtarget in {#3} {
    \expandafter\pcg@armtarget\pcgtarget()\pcg@end
    \ifx\pcg@tgtcol\@empty\else
      \pgfpointxy{\pcg@tgtcol}{0}\ifdim\pgf@x>\pcg@maxlanex \global\pcg@maxlanex=\pgf@x\fi\fi}%
  \edef\pcg@add{\noexpand\pcg@doarms{#2}{\pcg@drop}{\unexpanded{#3}}{\pcg@armlabels}}%
  \expandafter\g@addto@macro\expandafter\pcg@deferred\expandafter{\pcg@add}}
\def\pcg@armtarget#1(#2)#3\pcg@end{%
  \def\pcg@a{#1}\def\pcg@tgtcol{}%
  \ifx\pcg@a\pcg@retname
    \ifx\relax#2\relax \def\pcg@tgtc{(pcgexit |- \pcg@fn.south)}%
    \else \def\pcg@tgtc{(#2,0 |- \pcg@fn.south)}\def\pcg@tgtcol{#2}\fi
  \else \def\pcg@tgtc{(#1)}\fi}
\def\pcg@beside{beside}
\def\pcg@doarms#1#2#3#4{%
  \def\pcg@al{#4}%
  \foreach \pcgcase/\pcgtarget [count=\pcgi] in {#3} {\xdef\pcgarity{\pcgi}}%
  \foreach \pcgcase/\pcgtarget [count=\pcgi] in {#3} {%
    \pgfmathsetlengthmacro{\pcgdrop}{#2 + \pcg@armstep*((\pcgarity-1)/2 - abs(\pcgi-(\pcgarity+1)/2))}%
    \coordinate (pcgexit) at ($(#1.south west)!\pcgi/(\pcgarity+1)!(#1.south east)$);
    \expandafter\pcg@armtarget\pcgtarget()\pcg@end
    \path \pcg@tgtc coordinate (pcgtgt);
    \pgfpointdiff{\pgfpointanchor{pcgexit}{center}}{\pgfpointanchor{pcgtgt}{center}}%
    \pcg@dima=\pgf@x \ifdim\pcg@dima<0pt \pcg@dima=-\pcg@dima\fi 
    \pcg@measure{\pcg@dimb}{\tiny\ttfamily\pcgcase}\advance\pcg@dimb by 2mm 
    \ifnum\pcgi=1
      \ifdim\pcg@dima>\pcg@dimb \tikzset{pcg arm/.style={pos=0.25, above=1pt}}%
      \else \tikzset{pcg arm/.style={pos=0.75, left}}\fi
    \else\ifnum\pcgi=\pcgarity
      \ifdim\pcg@dima>\pcg@dimb \tikzset{pcg arm/.style={pos=0.25, above=1pt}}%
      \else \tikzset{pcg arm/.style={pos=0.75, right}}\fi
    \else
      \pgfmathparse{2*\pcgi<\pcgarity+1}%
      \ifnum\pgfmathresult=1 \tikzset{pcg arm/.style={pos=0.75, left}}%
      \else \tikzset{pcg arm/.style={pos=0.75, right}}\fi
    \fi\fi
    \draw[branch, rounded corners=2pt] (pcgexit) -- ++(0,-\pcgdrop) -| \pcg@tgtc
       node[guard, pcg arm] {\pcgcase};}}

\tikzset{name corner/.store in=\pcg@corner}   
\newcommand{\backdrop}[4][]{%
  \begingroup
  \tikzset{name corner=south west, #1}
  \begin{pgfonlayer}{pcgbackdrop}
    \edef\pcg@tmp{[backdrop, \unexpanded{#1}, fit={\unexpanded{#3}},
      label={[backdrop name, anchor=\pcg@corner]\pcg@corner:\unexpanded{#4}}] (#2)}%
    \expandafter\node\pcg@tmp {};
  \end{pgfonlayer}
  \endgroup}
\newcommand{\enclosure}[4][]{%
  \begingroup
  \tikzset{name corner=north west, #1}%
  \edef\pcg@tmp{[enclosure, \unexpanded{#1}, fit={\unexpanded{#3}},
    label={[enclosure name, anchor=\pcg@corner]\pcg@corner:\unexpanded{#4}}] (#2)}%
  \expandafter\node\pcg@tmp {};
  \endgroup}

\newcommand{\queue}[4][]{\pcg@node{queue}{#1}{#2}{#3}{#4}}
\newcommand{\feed}[4][]{%
  \pgfpointdiff{\pgfpointanchor{#2}{center}}{\pgfpointanchor{#3-entry}{center}}%
  \ifdim\pgf@x<1mm \ifdim\pgf@x>-1mm \def\pcg@start{#2.south}\else\def\pcg@start{#2}\fi\else\def\pcg@start{#2}\fi
  \draw[flow, rounded corners=2pt] (\pcg@start) -| (#3-entry) node[guard, pos=0.75, right, #1] {#4};}
\newcommand{\enter}[3][]{%
  \draw[flow] ([yshift=\pcg@enterlen]#2-entry) -- (#2-entry) node[guard, midway, right, #1] {#3};}
\makeatother

%% file: figures/lst-rtiow-render.tex
\begin{figure}[b]
\centering
 
\begin{lstlisting}[style=syscode, caption={A \emph{Ray Tracing in One Weekend} (RTIOW) path tracer~\cite{shirley2025rtw1} with a single sample per pixel, written in \sys.
% in destination-passing style~\cite{shaikhha2017dps}.
%\texttt{interact} performs dynamic dispatch on the material type.
ADT types for materials and interaction responses, and argument types, are omitted for brevity.}, label={lst:rtiow-render}]
func interact(ray, hit) $\shortrightarrow$ Response:
  match hit.material:
  | Lambertian(albedo) $\shortrightarrow$
    return Scatter{Ray{hit.pt, hit.n + rngu()}, albedo, 0}
  | Metal(albedo, fuzz) $\shortrightarrow$
    dir = reflect(ray.dir, hit.n) + fuzz * rngu()
    if dot(dir, hit.n) <= 0: return Absorb{0}  // grazing
    return Scatter{Ray{hit.pt, dir}, albedo, 0}
  | Dielectric(eta) $\shortrightarrow$ // reflect or refract
    return Scatter{Ray{hit.pt, ref(ray.dir, hit, eta)}, 1, 0}
  | DiffuseLight(tex) $\shortrightarrow$ return Absorb{eval(tex, hit.pt)}

func Li(L, beta, ray, n, scene) $\shortrightarrow$ Value:
  if n == 0: return L
  else match trace(ray, scene): // Find first hit geometry in scene
  | None $\shortrightarrow$ return L + beta * background(ray)
  | Some(hit) $\shortrightarrow$ match interact(ray, hit):
    | Absorb(e) $\shortrightarrow$ return L + beta * e
    | Scatter(dir, a, e) $\shortrightarrow$ return Li(L + beta * e, beta * a, dir, n-1, scene)

func render(width, height, out : mut, depth, scene):
  parfor p in [0, width * height): // compute pixel p
    out[p] = Li(0.0, 1.0, camera_ray(p), depth, scene)
\end{lstlisting}
\vspace{-0.95\textfloatsep}
\end{figure}

%% file: figures/fig-pcg-rtiow.tex
\begin{figure}[t]
\centering
\begin{subfigure}[b]{0.48\linewidth}\centering
\begin{pcg}
\begin{fn}[height=38mm, entry=center]{li}{Li}
  \cond{depth}{0,0}{n == 0}
  \dispatch{tr}{0,1.05}{trace}{hit}
  \op[right=7mm of tr.two east]{bg}{}{background} %
  \dispatch{ia}{0,2.3}{interact}{response}
  \case[lane=1.55]{depth}{T}{ret}
  \case{depth}{F}{tr}
  \case{tr}{None}{bg}
  \case{tr}{Some}{ia}
  \flow{bg}{ret}
  \case[lane=0.5]{ia}{Absorb}{ret}
  \case{ia}{Scatter}{self}
\end{fn}
The\op[above=3mm of li.north, font=\tiny\ttfamily, minimum height=9pt, inner xsep=2pt]{cam}{}{camera\_ray}
\queue[above=2.5mm of cam]{qy}{}{p}
\flow{qy}{cam}
\feed{cam}{li}{}
\end{pcg}
  \caption{\code{render} and \code{Li} of Listing~\ref{lst:rtiow-render}; calls are not expanded.}
  \label{fig:pcg-rtiow-li}
\end{subfigure}\hfill
\begin{subfigure}[b]{0.48\linewidth}\centering
\begin{pcg}
\begin{fn}[height=38mm]{ia}{interact}
  \cond[minimum width=12mm]{mat}{0,0}{hit.material}
    \op[rectangle split, rectangle split parts=4, fill=none, rectangle split part fill={color4!75, color4!75, color4!75, color2},
      every four node part/.style={font=\tiny\ttfamily}, minimum width=12mm]{met}{-7.25mm,2.05}
     {reflect\nodepart{two}rngu\nodepart{three}dot\nodepart{four}d <= 0}
  \op[below left=0pt and 2.5mm of met.north west, minimum width=12mm]{lam}{}{rngu}
  \op[below right=0pt and 2.5mm of met.north east, minimum width=12mm]{die}{}{ref}
  \op[right=2.5mm of die, minimum width=12mm]{lt}{}{eval}
  \arms[drop=3mm]{mat}{Lambertian/lam, Metal/met, Dielectric/die, DiffuseLight/lt} %
  \arms[drop=1mm]{met}{T/ret, F/ret}
  \flow{lam}{ret}
  \flow{die}{ret}
  \flow{lt}{ret}
\end{fn}
\enter{ia}{interact(ray, hit)}
\end{pcg}
  \caption{\code{interact}: dynamic dispatch on material type.}
  \label{fig:pcg-rtiow-interact}
\end{subfigure}
\caption{Parallel dispatch graphs (PDGs) representing divergence in the RTIOW path tracer. Blue queues are parallel loops; teal boxes are (unexpanded) calls; green boxes with dashed arrows are dispatches (conditional control flow); solid arrows are unconditional control flow; arrows to the bottom are returns.}
\label{fig:pcg-rtiow}
\end{figure}

%% file: figures/fig-defer-bg.tex
\begin{figure}[t]
\begin{minipage}[b]{0.43\linewidth}\centering
\begin{pcg}
\begin{fn}[entry=center, symmetric]{li}{Li}
  \cond{depth}{0,0}{n == 0}
  \dispatch{tr}{0,1.05}{trace}{hit}
  \dispatch{ia}{0,2.3}{interact}{response}
  \case[lane=0.75]{depth}{T}{ret}
  \case{depth}{F}{tr}
  \case{tr}{Some}{ia}
  \case{ia}{Absorb}{ret}
  \case{ia}{Scatter}{self}
\end{fn}
\op[above=3mm of li.north, font=\tiny\ttfamily, minimum height=9pt, inner xsep=2pt]{cam}{}{camera\_ray}
\queue[above=2.5mm of cam]{qy}{}{p}
\flow{qy}{cam}
\feed{cam}{li}{}
\queue[below=3mm of li.south]{q}{}{b}
\op[below=2.5mm of q, font=\tiny\ttfamily, minimum height=9pt, inner xsep=2pt]{bg}{}{background}
\flow{q}{bg}
\draw[branch, rounded corners=2pt] (tr.two east) -| (0.5,0 |- q)
  node[guard, pos=0, anchor=south west, xshift=1mm] {None} -- (q.east);
\end{pcg}
  \caption{PDG with \inputcode{background} deferred.}
  \label{fig:pcg-rtiow-defer-bg}
\end{minipage}\hfill
\begin{minipage}[b]{0.56\linewidth}
\begin{lstlisting}[style=halfcode, language=input,
% caption={\Cref{lst:rtiow-render} with deferred execution of \inputcode{background} via dataflow into a global work queue, corresponding to \Cref{fig:pcg-rtiow-defer-bg}.},
label={lst:rtiow-render-defer-bg},
caption={\sys-generated code for \Cref{fig:pcg-rtiow-defer-bg}: \Cref{lst:rtiow-render} with \inputcode{background} deferred into a global work queue.},
escapeinside={(*@}{@*)}]
// Li must communicate whether work was deferred
type Result = | Done(v : Value) | Defer(i : Index);

func Li(L, beta, ray, n, scene, b : mut) $\shortrightarrow$ Result:
 if n == 0: return Done(L)
 else match trace(ray, scene):
 | None $\shortrightarrow$ return Defer(append(b, {L, beta, ray}))
 | Some(hit) $\shortrightarrow$ match interact(ray, hit):
  | Absorb(e) $\shortrightarrow$ return Done(L + beta * e)
  | Scatter(dir, a, e) $\shortrightarrow$
    return Li(L+beta*e, beta*a, dir, n-1, scene, b)

func render(width, height, out : mut, depth, scene):
 b = queue()
 parfor p in [0, width * height):
  ray = camera_ray(p)
  match Li(0.0, 1.0, ray, depth, scene, b):
  | Done(v) $\shortrightarrow$ out[p] = v
  | Defer(i) $\shortrightarrow$ b[i].p = p // finish continuation

 // Drain the queue
 parfor (L, beta, ray, p) in b:
  out[p] = L + beta * background(ray)
\end{lstlisting}
\end{minipage}
\end{figure}

%% file: figures/fig-pcg-wavefront.tex
\begin{figure}[t]
\begin{minipage}[b]{0.56\linewidth}\centering
\begin{pcg}
\queue{l}{-28mm,3.15}{l}
\queue{m}{-14mm,3.15}{m}
\queue{d}{0mm,3.15}{d}
\queue{D}{14mm,3.15}{D}
\queue{b}{30mm,3.15}{b}
\begin{fn}[entry=center, symmetric, pad left=8mm, pad top=1.5mm, pad bottom=3mm]{rg}{}
  \dispatch{tr}{0,0}{trace}{hit}
  \cond[minimum width=12mm]{mat}{0,0.95}{hit.material}
  \case{tr}{Some}{mat}
  \arms[drop=9.5mm]{mat}{/l, /m, /d, /D}
\end{fn}
\path let \p1=($(mat.south west)!0.2!(mat.south east)$), \p2=(l) in
  node[guard, above=1pt] at ({(\x1+\x2)/2}, {\y1-9.5mm}) {Lambertian};
\node[guard, right=0.5mm] at ([yshift=2.75mm]m.north) {Metal};
\node[guard, right=0.5mm] at ([yshift=2.75mm]d.north) {Dielectric};
\node[guard, right=0.5mm] at ([yshift=2.75mm]D.north) {DiffuseLight};
\draw[branch, rounded corners=2pt] (tr.two east) -| (b)
  node[guard, pos=0, anchor=south west, xshift=3.5mm] {None};
\queue[above=2.5mm of rg.north]{t}{}{t}
\op[above=4mm of t, font=\tiny\ttfamily, minimum height=9pt, inner xsep=2pt]{cam}{}{camera\_ray}
\queue[above=2mm of cam]{pin}{}{i}
\flow{pin}{cam}
\flow{cam}{t}
\feed{t}{rg}{}

{\tikzset{fn/.append style={inner sep=2pt}}
\begin{fn}[entry=center, pad left=2mm, pad right=2mm, pad top=1.5mm, pad bottom=2mm]{cl}{}
  \op[below=4.5mm of l, font=\tiny\ttfamily, inner xsep=1.5pt]{kl}{}{rngu}
  \cond{nl}{-28mm,5.1}{$n$}
  \flow{kl}{nl}
  \case{nl}{T}{ret}
\end{fn}

\begin{fn}[entry=center, pad left=2mm, pad right=2mm, pad top=1.5mm, pad bottom=2mm]{cm}{}
  \dispatch[below=4.5mm of m, yshift=-5pt, font=\tiny\ttfamily, inner xsep=1.5pt]{km}{}{metal}{resp}
  \cond{nm}{-14mm,5.1}{$n$}
  \case[label={swap}]{km}{Scatter}{nm}
  \case[lane=-0.46, label={xshift=-1.5mm}]{km}{Abs.}{ret}
  \case{nm}{T}{ret}
\end{fn}

\begin{fn}[entry=center, pad left=2mm, pad right=2mm, pad top=1.5mm, pad bottom=2mm]{cd}{}
  \op[below=4.5mm of d, font=\tiny\ttfamily, inner xsep=1.5pt]{kd}{}{ref}
  \cond{nd}{0mm,5.1}{$n$}
  \flow{kd}{nd}
  \case{nd}{T}{ret}
\end{fn}

\begin{fn}[entry=center, pad left=2mm, pad right=2mm, pad top=1.5mm, pad bottom=8mm]{cD}{}
  \op[below=4.5mm of D, font=\tiny\ttfamily, inner xsep=1.5pt]{kD}{}{eval}
\end{fn}

\begin{fn}[entry=center, pad left=1mm, pad right=1mm, pad top=1.5mm, pad bottom=2mm]{cb}{}
  \op[below=4.5mm of b, font=\tiny\ttfamily, inner xsep=1.5pt]{bg}{}{background}
\end{fn}
}

\feed{l}{cl}{}
\feed{m}{cm}{}
\feed{d}{cd}{}
\feed{D}{cD}{}
\feed{b}{cb}{}

\coordinate (lane) at (-36mm,0);
\coordinate (busm) at ($(cl.south)+(0,-2.3mm)$);
\coordinate (busd) at ($(cl.south)+(0,-4.3mm)$);
\draw[branch, rounded corners=2pt] (nl.west) -| (lane |- t)
  node[guard, pos=0, anchor=south east, xshift=-0.5mm] {F} -- (t.west);
\draw[branch, -, rounded corners=2pt] (nm.west) -- ++(-4.7mm,0)
  node[guard, pos=0, anchor=south east, xshift=-0.5mm] {F}
  |- (lane |- busm) -- ([yshift=0.7mm]lane |- nl);
\draw[branch, -, rounded corners=2pt] (nd.west) -- ++(-4.7mm,0)
  node[guard, pos=0, anchor=south east, xshift=-0.5mm] {F}
  |- (lane |- busd) -- ([yshift=0.7mm]lane |- busm);

\backdrop[inner sep=1mm, minimum width=12.5mm]{gin}{(pin)(cam)}{}
\backdrop[name corner=north west]{optix}{(t)(rg)([yshift=-2mm]rg.south west)}{OptiX}
\node[note, fill=black!15, left=1.5mm of tr.text west] {RTCore};
\node[note, fill=black!15, below=1.5mm of kD.south] {TexUnit};
\backdrop[inner sep=1mm, minimum width=12.5mm]{gl}{(l)(cl)}{}
\backdrop[inner sep=1mm, minimum width=12.5mm]{gm}{(m)(cm)}{}
\backdrop[inner sep=1mm, minimum width=12.5mm]{gd}{(d)(cd)}{}
\backdrop[inner sep=1mm, minimum width=12.5mm]{gD}{(D)(cD)}{}
\backdrop[inner sep=1mm, minimum width=12.5mm]{gb}{(b)(cb)}{}

\enclosure[inner sep=1.5mm]{band}{(current bounding box)}{}
\queue[above=5mm of pin]{pband}{}{band}
\draw[flow] (pband) -- (pband |- band.north);
\end{pcg}
  \caption{Wavefront PDG: the continuation of \inputcode{Li} splits hits into per-material queues; scattering materials push into the trace queue; hardware acceleration is labeled in gray, unlabeled gray backgrounds map to standard GPU kernels.}
  \label{fig:pcg-rtiow-wavefront}
\end{minipage}\hfill
\begin{minipage}[b]{0.42\linewidth}
\begin{lstlisting}[language=input, mathescape=true, basicstyle=\scriptsize\ttfamily, backgroundcolor=\color{codebackground}, frame=single, basewidth=0.5em, captionpos=b, belowskip=0pt, belowcaptionskip=0pt,
% caption={Wavefront version of \Cref{lst:rtiow-render} following PBRT's style~\cite{pbrt4}. Corresponds to the PDG in \Cref{fig:pcg-rtiow-wavefront}. \inputcode{parfor}s are on GPU threads.},
caption={\sys-generated code for \Cref{fig:pcg-rtiow-wavefront}: a wavefront version of \Cref{lst:rtiow-render} in the style of PBRT~\cite{pbrt4}. \inputcode{parfor}s are on GPU threads, and iteration over a queue resets the size to 0.},
label={lst:rtiow-wavefront}]
func render(width, height, out, depth, scene):
 for band in [0, (width*height+(TILE-1))/TILE):
  t, b, l, m, d, D = queue() $\times$ 6
  parfor i in [0, TILE):
   p = band * TILE + i
   if p < width * height: // assumes depth > 0
    append(t, {0, 1, camera_ray(p), depth, p})
  while !empty(t): // loop induced by back arrow
   parfor (L, beta, ray, n, p) in t: // OptiX
    match trace(ray, scene): // RTCore
    | None $\shortrightarrow$ append(b, {L, beta, ray, p})
    | Some(hit) $\shortrightarrow$ s = {L, beta, ray, hit, n, p}
       match hit.material: // material binning
       | Lambertian(_) $\shortrightarrow$ append(l, s)
       | Metal(_) $\shortrightarrow$ append(m, s)
       | Dielectric(_) $\shortrightarrow$ append(d, s)
       | DiffuseLight(_) $\shortrightarrow$ append(D, s)
   // interact, specialized per material (queue)
   parfor (L, beta, ray, hit, n, p) in l:
    dir, a = lambertian(ray, hit) // Scatter
    if n - 1 > 0:
      append(t, {L, beta*a, dir, n-1, p})
    else: out[p] = L
   parfor (L, beta, ray, hit, n, p) in m:
    match metal(ray, hit):
    | Absorb(e) $\shortrightarrow$ out[p] = L + beta * e
    | Scatter(dir, a, _) $\shortrightarrow$
      if n - 1 > 0:
        append(t, {L, beta*a, dir, n-1, p})
      else: out[p] = L
   parfor (L, beta, ray, hit, n, p) in d:
    dir, a = dielectric(ray, hit) // Scatter
    if n - 1 > 0:
      append(t, {L, beta*a, dir, n-1, p})
    else: out[p] = L
   parfor (L, beta, ray, hit, n, p) in D:
    e = light(ray, hit) // on tex. unit, Absorb
    out[p] = L + beta * e
   parfor (L, beta, ray, p) in b:
    out[p] = L + beta * background(ray)
\end{lstlisting}
\end{minipage}
\end{figure}

%% file: figures/fig-compilation.tex
%

\colorlet{inputcol}{color1}   
\colorlet{ircol}{color3}      
\colorlet{outcol}{color5}     

\begin{figure*}[b]
    \centering
\begin{tikzpicture}[
  box/.style     = {draw, rounded corners=3pt, align=center,
                    minimum width=1.3cm, minimum height=0.65cm, font=\small\bfseries},
  input/.style   = {box, fill=inputcol, minimum width=\inw},
  ir/.style      = {box, fill=ircol, minimum height=1.1cm},
  output/.style  = {box, fill=outcol, minimum width=1.3cm},
  group/.style   = {draw, rounded corners=4pt, fill=inputcol,
                    minimum width=\inw, minimum height=1.6cm},
  title/.style   = {font=\small\bfseries, inner sep=3pt, anchor=north},
  sub/.style     = {draw, rounded corners=2pt, fill=white, inner sep=2pt, fill=color0,
                    minimum width=\inw-0.3cm, minimum height=0.4cm, font=\footnotesize},
  arr/.style     = {-{Stealth[length=6pt, width=5pt]}, thick, rounded corners=4pt},
  lbl/.style     = {font=\scriptsize, inner sep=2pt}
]
  \def\inw{3.1cm}                                        
  \pgfmathsetlengthmacro\slack{\linewidth - \inw - 5.7cm - 2pt} 
  \pgfmathsetlengthmacro\xI{0.5*\inw}                    
  \pgfmathsetlengthmacro\xP{\inw + 0.8cm + 0.46*\slack}  
  \pgfmathsetlengthmacro\xL{\xP + 1.6cm + 0.54*\slack}   
  \pgfmathsetlengthmacro\xB{\linewidth - 0.65cm - 2pt}   

  \node[group] (ALG) at (\xI, 0cm) {};
  \node[title] at (ALG.north) {Algorithm};
  \node[sub]   (SO) at (\xI,  0.12cm) {Set Operations~\cite{root2026bonsai}};
  \node[sub]   (PA) at (\xI, -0.42cm) {Parallel Annotations \S \ref{sec:parallel-annotations}};

  \node[input] (SCH) at (\xI, -1.45cm) {Schedule {\footnotesize\mdseries\S \ref{sec:schedule-syntax}}};
  \node[input] (LAY) at (\xI, -2.35cm) {Layout{\footnotesize\mdseries~\cite{gyurgyik2026decoupling}}};

  \node[ir] (PDG)  at (\xP, -0.15cm) {PDGs\\[-1pt]{\footnotesize\mdseries\S \ref{sec:ssa-blocks}}};
  \node[ir] (LLVM) at (\xL, -0.15cm) {LLVM\\[-1pt]{\footnotesize\mdseries\cite{lattner2004llvm}}};

  \node[output] (CPU) at (\xB,  0.40cm) {CPU};
  \node[output] (GPU) at (\xB, -0.70cm) {GPU};

  \draw[arr] (SO.east) -- node[lbl, above] {\bonsai~\cite{root2026bonsai} + \S\ref{sec:lowering-to-pdgs}} (SO.east -| PDG.west);
  \draw[arr] (PA.east) -- node[lbl, above] {PDG Lowering \S\ref{sec:lowering-to-pdgs}} (PA.east -| PDG.west);

  \draw[arr, sharp corners] ($(PDG.south)+(0.3,0)$) arc[start angle=0, end angle=-180, radius=0.3];
  \coordinate (loop-bot) at ($(PDG.south)+(0,-0.3)$);
  \draw[arr] (SCH.east) -| node[lbl, above, pos=0.25] {Transforms \S \ref{sec:transforms}} (loop-bot);

  \draw[arr] (PDG.east) --
       node[lbl, above]      {Code Generation \S \ref{sec:codegen}}
       node[lbl, below] (LS) {Layout Specialization~\cite{gyurgyik2026decoupling}}
     (LLVM.west);
  \draw[arr] (LAY.east) -| (LS.south);

  \draw[arr] (LLVM.east) -- (CPU.west);
  \draw[arr] (LLVM.east) -- (GPU.west);
\end{tikzpicture}
    \caption{
    \sys compilation pipeline. Set operations and parallel annotations are lowered to PDGs, which the schedule rewrites. Code generation and layout specialization perform some vector instruction selection (among other lowering optimizations) before using LLVM to target CPUs and GPUs.}
    \label{fig:overview}
\end{figure*}

%% file: figures/grammar-frontend.tex
\begin{wrapfigure}{r}{0.48\textwidth}
\vspace{-3em}
\footnotesize
\begin{tabular}{@{}l@{}}
\hline\\[-0.8em]
$D$ : Decl \;::=\; \code{func} $f$\code{(}$\overline{x\,\code{:}\,T}$\code{)} $\shortrightarrow$ $T$ \code{\{} $S^*$ \code{\}} \\
\phantom{$D$ : }\;|\; \code{type} $E$ \code{=} (\code{|} $V$\code{(}$\overline{x\,\code{:}\,T}$\code{)})$^+$\code{;} \hfill \algcomment{ADTs} \\[0.3em]
$T$ : Type \;::=\; $t$ \;|\; \code{mut} $t$ \;|\; \code{reduce(}$\oplus$\code{)} $t$ \\[0.3em] 
$S$ : Stmt \;::=\; $x$ \code{:} $T$ \code{=} $e$\code{;} \\
\phantom{$S$ : }\;|\; [\code{atomic}] $x$ $\oplus$\code{=} $e$\code{;} \hfill \algcomment{reductions} \\
\phantom{$S$ : }\;|\; \code{parfor} $i$ \code{in} $e_{lo}$\code{:}$e_{hi}$[\code{:}$e_s$] \code{\{} $S^*$ \code{\}} \hfill \algcomment{independent work} \\
\phantom{$S$ : }\;|\; \code{for} $i$ \code{in} $e_{lo}$\code{:}$e_{hi}$[\code{:}$e_s$] \code{\{} $S^*$ \code{\}} \hfill \algcomment{sequential} \\
\phantom{$S$ : }\;|\; \code{if} $e_b$ \code{\{} $S^*$ \code{\}} [\code{else} \code{\{} $S^*$ \code{\}}] \hfill \algcomment{sugar for \code{match}} \\
\phantom{$S$ : }\;|\; \code{match} $e$ \code{\{} ( | $V$\code{(}$\overline{x}$\code{)} $\shortrightarrow$ $S$)$^+$ \code{\}} \\
\phantom{$S$ : }\;|\; \code{while} $e_b$ \code{\{} $S^*$ \code{\}} \\
\phantom{$S$ : }\;|\; \code{return} [$e$]\code{;} \;|\; $g$\code{(}$\overline{e}$\code{);} \\[0.3em]
$e$ : Expr, \ $t$ : type \;::=\; $\ldots$ \hfill \algcomment{\bonsai~\cite{root2026bonsai}} \\[0.3em]
\hline
\end{tabular}
\vspace{-0.5em}
\caption{\sys's extensions to \bonsai~\cite{root2026bonsai, gyurgyik2026decoupling}.}
\label{fig:frontend-grammar}
\vspace{-2em}
\end{wrapfigure}

%% file: figures/fig-schedule-lang.tex
\begin{figure*}[t]
\centering
\footnotesize
\newcommand{\schedrow}[2]{\makebox[\linewidth][s]{#1\hfill#2\hfil}}%
\ifdefined\schednamew\else\newlength{\schednamew}\fi\settowidth{\schednamew}{\schedule{specialize}}%
\newcommand{\sigrow}[2]{\makebox[\linewidth][s]{\makebox[\schednamew][l]{\schedule{#1}}\;:\;#2\hfil}}%
\begin{minipage}[t]{0.51\textwidth}
\begin{tabular}[t]{@{}l@{}}
\hline\\[-0.8em]
\schedrow{$S$ : Schedule \;::=\; \schedule{schedule \{} $D^*$ \schedule{\}}}{} \\
\schedrow{$D$ : Directive \;::=\; [$c$ (\schedule{,} $c$)$^*$ \schedule{=}] ($f$ \;|\; $q$) (\schedule{.}($L$ \;|\; $K$ \;|\; $H$))$^+$}{} \\
\schedrow{$c$ : Cursor \;::=\; $l$ \;|\; $f$ \;|\; $q$ \;|\; $e$}{\algcomment{handles into a PDG}} \\
\schedrow{$l$ : Loop cursor \;::=\; $i$ \;|\; $x$\schedule{::}$V$\schedule{.}$a$}{} \\
\schedrow{$e$ : Edge cursor \;::=\; $g$ \;|\; $g$\schedule{.cont} \;|\; $x$\schedule{::}$V$}{\algcomment{call, call return, arm}} \\
\schedrow{$o$ : Owner \;::=\; $l$ \;|\; \schedule{root}}{\algcomment{a loop, or loop nest root}} \\[0.3em]
\schedrow{$L$ : Loop \;::=\; \schedule{split(}$l$\schedule{,} $n$\schedule{)}}{\algcomment{tiling loop $l$ by a factor of $n$}} \\
\schedrow{\phantom{$L$ : }\;|\; \schedule{collapse(}$l_o$\schedule{,} $l_i$\schedule{)}}{\algcomment{loop flattening}} \\
\schedrow{\phantom{$L$ : }\;|\; \schedule{reorder(}$l_i$\schedule{,} $l_o$\schedule{)}}{\algcomment{loop interchange}} \\
\schedrow{\phantom{$L$ : }\;|\; \schedule{persistent(}$l$\schedule{,} $n$\schedule{)}}{\algcomment{$n$ iterations claim $l$ iterations}} \\
\schedrow{\phantom{$L$ : }\;|\; \schedule{unroll(}$l$\schedule{)}}{\algcomment{loop unrolling}} \\
\schedrow{\phantom{$L$ : }\;|\; \schedule{vectorize(}$l$\schedule{)}}{\algcomment{vector lane per iteration}} \\[0.3em]
\schedrow{$K$ : Control \;::=\; \schedule{queue(}$o$ [\schedule{,} $M$] [\schedule{,} $n$]\schedule{)}}{\algcomment{queue owned at $o$}} \\
\schedrow{\phantom{$K$ : }\;|\; \schedule{defer(}$e$\schedule{,} $q$\schedule{)}}{\algcomment{continuation at $e$ onto $q$}} \\
\schedrow{\phantom{$K$ : }\;|\; \schedule{specialize(}$c$\schedule{,} $x$\schedule{)}}{\algcomment{execution per variant of $x$}} \\
\schedrow{\phantom{$K$ : }\;|\; \schedule{sort(}$c$\schedule{,} $\lambda$\schedule{)}}{\algcomment{instances of $c$ ordered by $\lambda$}} \\
\schedrow{\phantom{$K$ : }\;|\; \schedule{loopify(}[$n$] [\schedule{,} $M$]\schedule{)}}{\algcomment{recursion to a loop; stack $n$}} \\[0.3em]
\schedrow{$H$ : Hardware \;::=\; \schedule{bind(}$c$\schedule{,} $R$\schedule{)}}{\algcomment{one $R$ per instance of $c$}} \\
\schedrow{\phantom{$H$ : }\;|\; \schedule{prefetch(}$l$ [\schedule{,} $n$]\schedule{)}}{\algcomment{$n$ cache lines prefetched}} \\[0.3em]
\hline
\end{tabular}
\end{minipage}\hfill\vrule\hfill
\begin{minipage}[t]{0.45\textwidth}
\begin{tabular}[t]{@{}l@{}}
\hline\\[-0.8em]
\schedrow{$\mathit{Args}, \mathit{Key}, \mathit{Var}$}{\algcomment{Instance arguments, sort key, variable}} \\
\schedrow{$\mathit{Resource}, \mathit{Storage}$}{\algcomment{Machine-dependent, \Cref{sec:hardware-models}}} \\
\schedrow{$\mathit{Cursor} = \mathit{Node} \mid \mathit{Edge}$}{\algcomment{Handle into a PDG}} \\
\schedrow{$\mathit{Node} = \mathit{Loop} \mid \mathit{Func}$}{\algcomment{Loops and calls}} \\
\schedrow{$\mathit{Queue} \mathrel{<:} \mathit{Loop}$}{\algcomment{Queue drain is always a loop}} \\
\schedrow{$\mathit{Owner} = \mathit{Loop} \mid \schedule{root}$}{\algcomment{Queue owners}} \\[0.3em]
\sigrow{split}{$\mathit{Loop} \times \mathit{Int} \to \mathit{Loop} \times \mathit{Loop}$} \\
\sigrow{collapse}{$\mathit{Loop} \times \mathit{Loop} \to \mathit{Loop}$} \\
\sigrow{reorder}{$\mathit{Loop} \times \mathit{Loop} \to ()$} \\
\sigrow{persistent}{$\mathit{Loop} \times \mathit{Int} \to \mathit{Loop}$} \\
\sigrow{unroll}{$\mathit{Loop} \to ()$} \\
\sigrow{vectorize}{$\mathit{Loop} \to ()$} \\[0.3em]
\sigrow{queue}{$\mathit{Owner}\; [\times \mathit{Storage}]\; [\times \mathit{Int}] \to \mathit{Queue}$} \\
\sigrow{defer}{$\mathit{Edge} \times \mathit{Queue} \to \mathit{Func}$} \\
\sigrow{specialize}{$\mathit{Node} \times \mathit{Var} \to \mathit{Node} \times \cdots \times \mathit{Node}$} \\
\sigrow{sort}{$\mathit{Node} \times (\mathit{Args} \to \mathit{Key}) \to ()$} \\
\sigrow{loopify}{$[\mathit{Int}] [\times \mathit{Storage}] \to ()$} \\[0.3em]
\sigrow{bind}{$\mathit{Node} \times \mathit{Resource} \to ()$} \\
\sigrow{prefetch}{$\mathit{Loop}\; [\times \mathit{Int}] \to ()$} \\[0.3em]
\hline
\end{tabular}
\end{minipage}
\par\smallskip
\begin{subfigure}[t]{0.51\textwidth}
\caption{Syntax. $f$, $g$ name functions, $q$ a queue, $i$ a loop, $x$ a variable, $V$ a variant, $a$ a field, $n$ an integer, and $R$, $M$ a resource and storage kind (\Cref{sec:hardware-models}); $x$\schedule{::}$V$\schedule{.}$a$ is a data-structure-induced loop over $V$'s field $a$.
}
\label{fig:schedule-syntax}
\end{subfigure}\hfill
\begin{subfigure}[t]{0.45\textwidth}
\caption{Signatures. Each transform is a method of the cursor it is applied within: a function, or a queue for its drain. Result types are the cursors produced by the transform.}
\label{fig:schedule-signatures}
\end{subfigure}
\caption{\sys's scheduling language. Directives apply in written order. [...] marks an optional argument/type.}
\label{fig:schedule-grammar}
\end{figure*}

%% file: figures/fig-hardware-models.tex
\begin{wrapfigure}{r}{0.42\columnwidth}
\vspace{-1.25em}
\scriptsize
\begin{tikzpicture}[x=1em, y=-1.15em,
  every node/.style={inner sep=1pt, outer sep=0pt},
  edge/.style={line width=0.4pt}]
\node[anchor=west] at (0,0) {\textit{CPU}};
\node[anchor=west] at (0,1) {host};
\node[anchor=east] at (\linewidth,1) {\algcomment{\schedule{Heap} \schedule{Stack}}};
\node[anchor=west] at (1,2) {\schedule{CPUThread}};
\node[anchor=east] at (\linewidth,2) {\algcomment{\schedule{Stack} \schedule{ThreadLocal}}};
\draw[edge] (0.4,1.45) -- (0.4,2) -- (0.9,2);
\node[anchor=west] at (0,3.6) {\textit{GPU}};
\node[anchor=west] at (0,4.6) {host};
\node[anchor=east] at (\linewidth,4.6) {\algcomment{\schedule{Device} \schedule{Managed}}};
\node[anchor=west] at (1,5.6) {\schedule{GPUBlock}};
\node[anchor=east] at (\linewidth,5.6) {\algcomment{\schedule{DeviceShared}}};
\node[anchor=west] at (2,6.6) {\schedule{GPUThread}};
\node[anchor=east] at (\linewidth,6.6) {\algcomment{\schedule{Stack}}};
\node[anchor=west] at (3,7.6) {\schedule{TextureUnit}};
\node[anchor=west] at (1,8.6) {\schedule{OptixThread}};
\node[anchor=east] at (\linewidth,8.6) {\algcomment{\schedule{Stack}}};
\node[anchor=west] at (2,9.6) {\schedule{RTCore}};
\node[anchor=west] at (2,10.6) {\schedule{TextureUnit}};
\draw[edge] (0.4,5.05) -- (0.4,8.6) -- (0.9,8.6);
\draw[edge] (0.4,5.6) -- (0.9,5.6);
\draw[edge] (1.4,6.05) -- (1.4,6.6) -- (1.9,6.6);
\draw[edge] (2.4,7.05) -- (2.4,7.6) -- (2.9,7.6);
\draw[edge] (1.4,9.05) -- (1.4,10.6) -- (1.9,10.6);
\draw[edge] (1.4,9.6) -- (1.9,9.6);
\end{tikzpicture}
\caption{
\sys's two machine models: compute resources $R$ in their hierarchy and storage kinds $M$ at the level that owns them.
}
\label{fig:hardware-models}
\vspace{-2em}
\end{wrapfigure}

%% file: figures/lst-rtiow-ir.tex
\begin{figure}[H]
    \centering
\begingroup
\setlength{\multicolsep}{0pt}
\captionof{lstlisting}{\Cref{lst:rtiow-render}'s path tracer lowered to \sys's IR. \irlang|ret|s are inlined and types are omitted for brevity.}
\label{lst:rtiow-ir}
\begin{multicols}{2}
\begin{lstlisting}[language=ir, mathescape=true, basicstyle=\scriptsize\ttfamily, backgroundcolor=\color{codebackground}, frame=single, aboveskip=0pt, belowskip=0pt]
func interact(ray, hit):
 m = hit.material
 cjmp m [lamb(ray, hit), metal(ray, hit),
         diel(ray, hit), light(hit)]
lamb(albedo, ray, hit):
 call rngu() lamb_c(albedo, hit)
lamb_c(u, albedo, hit):
 d = hit.n + u
 jmp scatter(d, albedo, hit)
metal(albedo, fuzz, ray, hit):
 call reflect(ray.dir, hit.n)
      metal_c(albedo, fuzz, hit)
metal_c(rd, albedo, fuzz, hit):
 call rngu() metal_cc(rd, albedo, fuzz, hit)
metal_cc(u, rd, albedo, fuzz, hit):
 fu = fuzz * u
 d = rd + fu
 call dot(d, hit.n) metal_ccc(albedo, hit, d)
metal_ccc(c, albedo, hit, d):
 g = c <= 0
 cjmp g [scatter(d, albedo, hit), // false
         metal_a()]               // true
metal_a(): // grazing
 a = Absorb{0}
 ret a
diel(eta, ray, hit):
 call ref(ray.dir, hit, eta) scatter(1, hit)
light(tex, hit):
 call eval(tex, hit.pt) light_c()
light_c(e):
 a = Absorb{e}
 ret a

// jmp target (lamb_c), cjmp arm (metal_ccc),
// and call continuation (diel)
scatter(d, albedo, hit):
 r = Ray{hit.pt, d}
 s = Scatter{r, albedo, 0}
 ret s
\end{lstlisting}
\columnbreak
\begin{lstlisting}[language=ir, mathescape=true, basicstyle=\scriptsize\ttfamily, backgroundcolor=\color{codebackground}, frame=single, aboveskip=0pt, belowskip=0pt, firstnumber=last]
func Li(L, beta, ray, n, scene):
 c = n == 0
 cjmp c [li_tr(L, beta, ray, n, scene), ret L]
li_tr(L, beta, ray, n, scene):
 call trace(ray, scene)
      li_hit(L, beta, ray, n, scene)
li_hit(h, L, beta, ray, n, scene):
 cjmp h [li_miss(L, beta, ray),
         li_some(L, beta, ray, n, scene)]
li_miss(L, beta, ray):
 call background(ray) li_emit(L, beta)
li_some(hit, L, beta, ray, n, scene):
 call interact(ray, hit)
      li_resp(L, beta, n, scene)
li_resp(resp, L, beta, n, scene):
 cjmp resp [li_emit(L, beta),
            li_scatter(L, beta, n, scene)]
li_emit(e, L, beta): // miss and Absorb
 t = beta * e
 r = L + t
 ret r
li_scatter(dir, a, e, L, beta, n, scene):
 t = beta * e
 L1 = L + t
 beta1 = beta * a
 n1 = n - 1
 call Li(L1, beta1, dir, n1, scene) ret

func render(width, height, out, depth, scene):
 N = width * height
 par p [0, N, 1] body(out, depth, scene) ret
body(p, out, depth, scene):
 call camera_ray(p) cont(p, out, depth, scene)
cont(r, p, out, depth, scene):
 call Li(0.0, 1.0, r, depth, scene)
      body_store(p, out)
body_store(v, p, out):
 store out[p], v
 yield
\end{lstlisting}
\end{multicols}
\endgroup
\end{figure}

%% file: figures/alg-defer.tex
\begin{algorithm}[H]
\caption{Defer Implementation, shown for a call edge \irlang|call g(a) k(b)|. For a call continuation the stored state is the return value \irlang|v| with \irlang|b|, and the drain resumes at \irlang|k(v, b)|; for a dispatch arm it is the arm's arguments, and the drain resumes at the arm. $\mathit{func}(q)$ is the function containing the loop that owns $q$; $w$ is the stored state, which the drain $D$ passes to every block (elided); \irlang|Defer_q| is the \irlang|Defer| arm for queue $q$.}
\label{alg:defer}
\scriptsize
\def\kw#1{\textnormal{\textbf{#1}}}
\makeatletter\def\Cont{\Statex\hskip\ALG@tlm\hspace{2em}}\makeatother
\lstset{basicstyle=\ttfamily\scriptsize}
\renewcommand{\algorithmicindent}{1em}
\algrenewcommand{\alglinenumber}[1]{\scriptsize #1:}
\setlength{\multicolsep}{0pt}
\setlength{\columnsep}{1em}
\begin{multicols}{2}
\begin{algorithmic}[1]
\algnotext{EndFor}\algnotext{EndIf}\algnotext{EndFunction}
\Function{Defer}{$f, g, q$} \Comment{$c$ = \schedule{f.defer(g, q)}}
\ForAll{$s =$ \irlang|call g(a) k(b)| $\in f$}
  \State \kw{rewrite} $s$ \kw{to} \irlang|defer (s, a ++ b) q|
  \State $q.\mathit{producers} \mathrel{+}= s$
\EndFor
\State \Call{Propagate}{$f, q$}
\State \Return \Call{EmitDrain}{$q$} \Comment{the cursor $c$}
\EndFunction
\Statex
\Function{Propagate}{$f, q$} \Comment{up to $q$'s loop}
\If{$f = \mathit{func}(q)$} \Return \Comment{loop body yields} \EndIf
\If{$\mathit{callers}(f) = \emptyset$} \kw{error} \Comment{$g$ not under $q$'s owner} \EndIf
\If{\irlang|Defer_q| $\in \mathit{rettype}(f)$} \Return \EndIf
\If{$\mathit{rettype}(f) \neq$ \irlang|Result|}
  \State $\mathit{rettype}(f) \gets$ \irlang|Result| \Comment{\irlang|Done(T)| $\mid$ \irlang|Defer_*(Index)|}
  \State \kw{rewrite each} \irlang|ret v| \kw{to} \irlang|ret Done(v)|
\EndIf
\State $\mathit{rettype}(f) \mathrel{+}=$ \irlang|Defer_q(Index)| \Comment{one arm per queue}
\ForAll{$s' =$ \irlang|call f(a) k(b)| $\in \mathit{callers}(f)$}
  \If{$k \neq$ \irlang|ret|} \Call{EmitPatch}{$s', q$} \EndIf
  \State \Call{Propagate}{$\mathit{func}(s'), q$}
\EndFor
\EndFunction
\Statex
\Function{EmitPatch}{$s', q$} \Comment{$s' =$ \irlang|call f(a) k(b)|}
\State \kw{assert} $\mathit{func}(s') \notin \mathit{callees}^{*}(f)$ \Comment{non-tail recursion}
\State \kw{rewrite} $s'$ \kw{to} \irlang|call f(a) k'(b)|
\State \kw{emit} \irlang|k'(x, b):| \Comment{$x$ : \irlang|Result|}
\Cont \irlang|cjmp x [k(b), p_q(b), ...]| \Comment{\irlang|Done|$\shortrightarrow$\irlang|k|, \irlang|Defer_q|$\shortrightarrow$\irlang|p_q|}
\State \kw{emit} \irlang|p_q(i, b):| \Comment{save this frame}
\Cont \irlang|q[i].frames += (s', b)|
\Cont \irlang|ret Defer_q(i)| \Comment{\irlang|yield| in $q$'s loop body}
\State $q.\mathit{patches} \mathrel{+}= (s', k)$
\EndFunction
\columnbreak
\Function{EmitDrain}{$q$} \Comment{rebuilt per directive}
\State \kw{emit} \irlang|D(w):| \Comment{an arm per $s \in q.\mathit{producers}$}
\Cont \irlang|cjmp w.tag [L_s(w), ...]|
\ForAll{$s =$ \irlang|call g(a) k(b)| $\in q.\mathit{producers}$}
  \State \kw{emit} \irlang|L_s(w): call g(w.a) k(w.b)| \Comment{resume the call}
  \State \Call{EmitCopy}{$k$}
\EndFor
\ForAll{$(s', k) \in q.\mathit{patches}$} \Call{EmitCopy}{$k$} \EndFor
\State \Call{EmitReturn}{$q$}
\State $q.\mathit{drain} \gets$ \kw{par} $w \in q$ \kw{do} $D(w)$
\State \Call{PlaceAtOwner}{$q$}
\State \Return $D$
\EndFunction
\Statex
\Function{EmitCopy}{$k$} \Comment{once per $k$}
\State $\mathit{blocks}(D) \mathrel{\cup}= \mathit{copy}(\mathit{reach}(k))$ \Comment{$k$ and its successors}
\State \kw{rewrite each} \irlang|ret x| \kw{in the copy to} \irlang|jmp Return(x)|
\EndFunction
\Statex
\Function{EmitReturn}{$q$} \Comment{returning inside $D$}
\State \kw{emit} \irlang|Return(x):| \Comment{$x$ : \irlang|Result|}
\Cont \irlang|cjmp x [Resume(), Forward_q'(), ...]| \Comment{per queue $q'$}
\State \kw{emit} \irlang|Resume(v):| \Comment{innermost frame first}
\Cont \irlang|(t, b) = pop_front(w.frames)|
\Cont \irlang|cjmp t [k(v, b), ...]| \Comment{$(s', k) \in q.\mathit{patches}$}
\State \kw{emit} \irlang|Forward_q'(i):| \Comment{deferred again, into $q'$}
\Cont \irlang|q'[i].frames ++= w.frames; yield|
\EndFunction
\Statex
\Function{PlaceAtOwner}{$q$} \Comment{legality, \Cref{sec:transforms}}
\State \kw{insert} $q.\mathit{drain}$ \kw{at} $\mathit{owner}(q)$ \kw{after every producer}
\ForAll{$S$ between producer $P$ and $q.\mathit{drain}$, $d \in \mathit{data}(D)$}
  \State \kw{assert} $\mathit{access}_S(d) = \mathit{access}_P(d)$
\EndFor
\If{$D$ feeds a queue upstream} \kw{loop until empty} \EndIf
\EndFunction
\end{algorithmic}
\end{multicols}
\end{algorithm}

%% file: figures/fig-pcg-pbrt.tex
\newcommand{\kop}[3]{\op[below=2.5mm of #1, font=\tiny\ttfamily, minimum height=9pt, inner xsep=2pt]{#2}{}{#3}}
\newcommand{\yw}{4.8}
\newcommand{\yr}{5.1}
\newcommand{\xw}{76mm}
\newcommand{\xr}{79mm}
\newcommand{\drops}[1]{%
  \coordinate (#1-wf) at ($(#1.south)+(1.3mm,0)$);
  \coordinate (#1-rf) at ($(#1.south)+(-1.3mm,0)$);
  \draw[flow,-] (#1-wf) -- (#1-wf |- 0,\yw);
  \draw[flow,-,sharp corners] (#1-rf) -- ($(#1-rf |- 0,\yw)+(0,0.6mm)$)
    arc[start angle=90, end angle=-90, radius=0.6mm] -- (#1-rf |- 0,\yr);}

\begin{figure}[H]
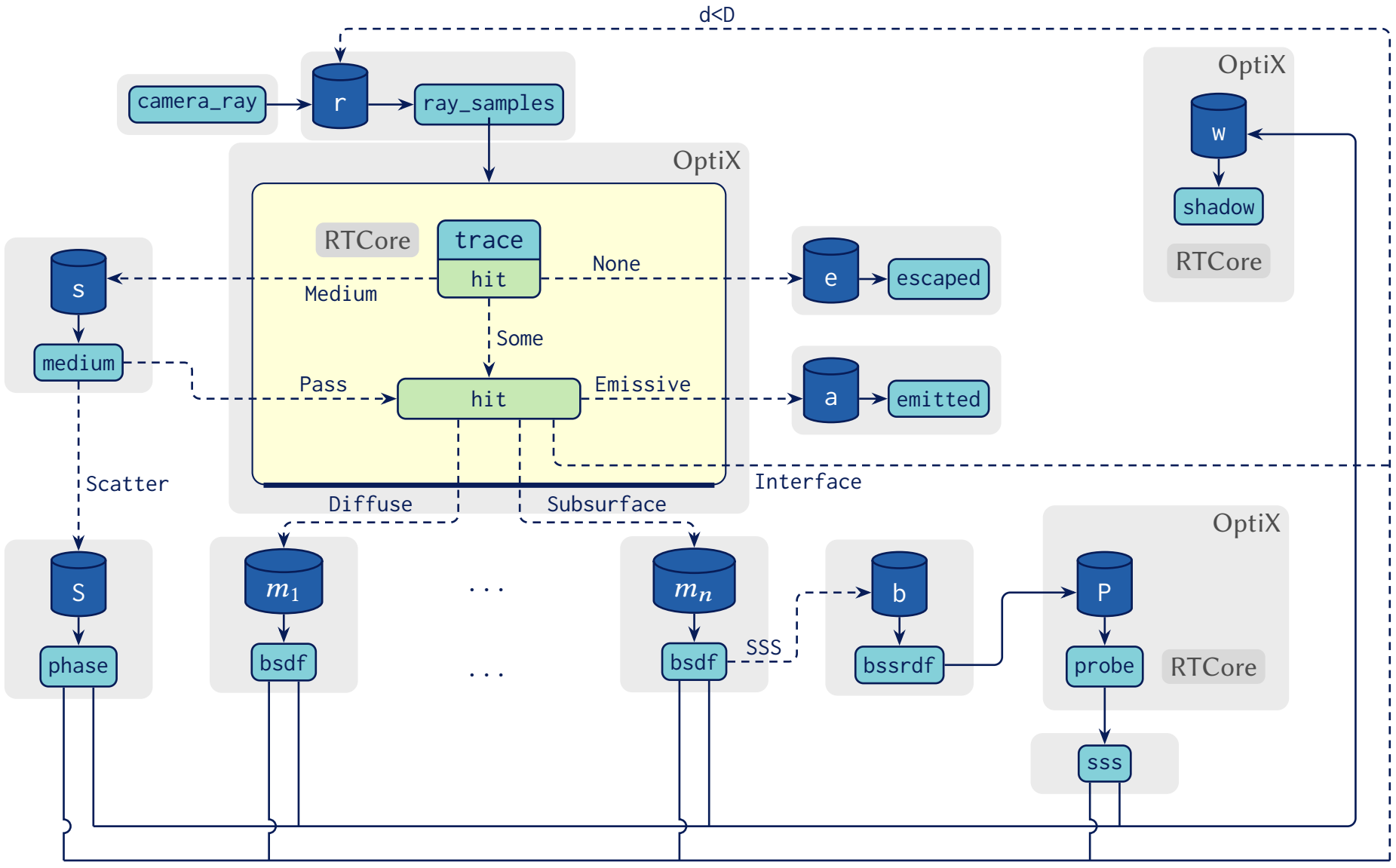

\centering
\begin{pcg}
\queue{s}{-36mm,0.1}{s}
\queue{e}{30mm,0}{e}
\queue{a}{30mm,1.05}{a}
\queue{m1}{-18mm,2.75}{$m_1$}
\queue{mn}{18mm,2.75}{$m_n$}
\node[text=color8] at (0mm,2.75) {$\cdots$};
\node[text=color8] at (0mm,3.5) {$\cdots$};
\begin{fn}[entry=center, symmetric, pad left=11mm, pad top=1.5mm, pad bottom=4mm]{rg}{}
  \dispatch{tr}{0,0}{trace}{hit}
  \cond[minimum width=16mm]{hit}{0,1.05}{hit}
  \case{tr}{Some}{hit}
  \case[label={pos=0.29}]{tr}{None}{e}
  \case[label={pos=0.29, below}]{tr}{Medium}{s}
  \case[label={pos=0.28}]{hit}{Emissive}{a}
  \arms[drop=9mm]{hit}{Diffuse/m1, Subsurface/mn}
\end{fn}
\coordinate (ifc) at ($(hit.south)+(0,-4mm)$);
\draw[branch,-, rounded corners=2pt] ($(hit.south west)!0.85!(hit.south east)$) -- ++(0,-4mm) -- (\xr,0 |- ifc);
\node[guard, below] at (28mm,0 |- ifc) {Interface};
\op[right=2.5mm of e, font=\tiny\ttfamily, minimum height=9pt, inner xsep=2pt]{ke}{}{escaped}
\flow{e}{ke}
\op[right=2.5mm of a, font=\tiny\ttfamily, minimum height=9pt, inner xsep=2pt]{ka}{}{emitted}
\flow{a}{ka}
\kop{s}{ks}{medium}
\flow{s}{ks}
\queue{S}{-36mm,2.75}{S}
\case[label={pos=0.6}]{ks}{Scatter}{S}
\draw[branch, rounded corners=2pt] (ks.east) -- ++(6mm,0) |- (hit.west);
\node[guard, above] at (-14.5mm,0 |- hit.west) {Pass};
\op[above=5mm of rg.north, font=\tiny\ttfamily, minimum height=9pt, inner xsep=2pt]{smp}{}{ray\_samples}
\queue[left=4mm of smp]{r}{}{r}
\op[left=4mm of r, font=\tiny\ttfamily, minimum height=9pt, inner xsep=2pt]{cam}{}{camera\_ray}
\flow{cam}{r}
\flow{r}{smp}
\feed{smp}{rg}{}
\kop{S}{kS}{phase}  \flow{S}{kS}
\kop{m1}{k1}{bsdf}  \flow{m1}{k1}
\kop{mn}{kn}{bsdf}  \flow{mn}{kn}
\queue{b}{36mm,2.75}{b}
\draw[branch, rounded corners=2pt] (kn.east) -- (27mm,0 |- kn)
  node[guard, pos=0, anchor=south west, xshift=1mm] {SSS} |- (b.west);
\kop{b}{bk}{bssrdf} \flow{b}{bk}
\queue{P}{54mm,2.75}{P}
\draw[flow, rounded corners=2pt] (bk.east) -- (45mm,0 |- bk) |- (P.west);
\kop{P}{pr}{probe} \flow{P}{pr}
\node[note, fill=black!15, right=1.5mm of pr] (rt3) {RTCore};
\op[below=5mm of pr, font=\tiny\ttfamily, minimum height=9pt, inner xsep=2pt]{sx}{}{sss} \flow{pr}{sx}
\drops{kS} \drops{k1} \drops{kn} \drops{sx}
\queue{w}{64mm,-1.27}{w}
\draw[flow, rounded corners=2pt] (kS-wf |- 0,\yw) -- (\xw,\yw) |- (w.east);
\kop{w}{sh}{shadow} \flow{w}{sh}
\node[note, fill=black!15, below=1.5mm of sh] (rt2) {RTCore};
\draw[flow,-] (kS-rf |- 0,\yr) -- (\xr,\yr);
\coordinate (lt) at ($(r.north)+(0,3mm)$);
\draw[branch, rounded corners=2pt] (\xr,\yr) -- (\xr,0 |- lt) -| (r.north);
\node[guard, above] at (20mm,0 |- lt) {d<D};
\backdrop[inner sep=1mm, minimum width=13mm]{gcam}{(cam)}{}
\backdrop[inner sep=1mm, minimum width=13mm]{gsmp}{(r)(smp)}{}
\backdrop[name corner=north east]{optix}{(rg)([yshift=-0.5mm]rg.south west)([yshift=1.5mm]rg.north east)}{OptiX}
\node[note, fill=black!15, left=1.5mm of tr.text west] {RTCore};
\backdrop[inner sep=1mm, minimum width=13mm]{gs}{(s)(ks)}{}
\backdrop[inner sep=1mm, minimum width=13mm]{ge}{(e)(ke)}{}
\backdrop[inner sep=1mm, minimum width=13mm]{ga}{(a)(ka)}{}
\backdrop[inner sep=1mm, minimum width=13mm]{gS}{(S)(kS)}{}
\backdrop[inner sep=1mm, minimum width=13mm]{g1}{(m1)(k1)}{}
\backdrop[inner sep=1mm, minimum width=13mm]{gn}{(mn)(kn)}{}
\backdrop[inner sep=1mm, minimum width=13mm]{gb}{(b)(bk)}{}
\backdrop[name corner=north east]{optix3}{(P)(pr)(rt3)([yshift=2mm]P.north)}{OptiX}
\backdrop[inner sep=1mm, minimum width=13mm]{gx}{(sx)}{}
\backdrop[name corner=north east]{optix2}{(w)(sh)(rt2)([yshift=2mm]w.north)}{OptiX}
\end{pcg}
\caption{Inner tile (with \code{band} and \code{i} nested loops) of PBRT's GPU path integrator~\cite[Ch.~15.2]{pbrt4} expressed as a PDG. Surface hits are binned into per-material queues (two shown), so every \inputcode{bsdf} kernel is specialized to a material. \inputcode{r} is one logical queue that PBRT double-buffers, as it has a self-edge (its own drain can push to it).}
\label{fig:pcg-pbrt-wavefront}
\end{figure}